\documentclass[iop]{emulateapj}                                                                                                                             
\usepackage{apjfonts}
\usepackage{graphicx}
\usepackage{amsmath}
\usepackage{amssymb}
\usepackage{multirow}
\usepackage{rotating}
\usepackage{booktabs}
\usepackage{times}
\usepackage{ulem}

\newcommand{\zmag}{\ensuremath{z^\prime}}
\newcommand{\Msun}{\ensuremath{M_{\odot}}}
\newcommand{\Zsun}{\ensuremath{Z_{\odot}}}
\newcommand{\lya}{Ly$\alpha$}

\newcommand{\ha}{H$\alpha$}

\slugcomment{Accepted for publication in ApJ}

\shorttitle{Stellar populations of LAEs at $z=4.86$}
\shortauthors{Yuma et al.}
\begin{document}
\title{Stellar Populations of Lyman-alpha Emitters at $z=4.86$: \\A Comparison to $z\sim5$ LBGs\altaffilmark{1}}
\author{Suraphong Yuma        \altaffilmark{2}, 
        Kouji Ohta        \altaffilmark{2},
        Kiyoto Yabe        \altaffilmark{2},
        Kazuhiro Shimasaku   \altaffilmark{3},
        Makiko Yoshida     \altaffilmark{3}, \\
        Masami Ouchi       \altaffilmark{4}, 
        Ikuru Iwata       \altaffilmark{5}, and
        Marcin Sawicki     \altaffilmark{6}
        }

\email{yuma@kusastro.kyoto-u.ac.jp}

\begin{abstract}
We present a study of stellar population of Lyman Alpha Emitters (LAEs) at $z = 4.86$ in the Great Observatories Origins Deep Survey (GOODS) North field and its flanking field. The LAEs are selected based on optical narrowband (NB711) and broadband ($V$, $I_c$, and \zmag) observations by Suprime-Cam attached on Subaru Telescope. With the publicly available IRAC data in GOODS-N and further IRAC observations in the flanking fields, we select five LAEs which are not contaminated by neighboring objects in IRAC images and construct their observed spectral energy distributions (SEDs) with $I_c$, $z^{\prime}$, IRAC 3.6\micron, and 4.5\micron~band photometry. The SEDs cover the rest-frame UV to optical wavelengths. We derive stellar masses, ages, color excesses, and star formation rates of five LAEs using SED fitting method. Assuming the constant star formation history, we find that the stellar masses  range from $10^8$ to $10^{10}$ $M_{\odot}$ with the median value of $2.5\times10^9\Msun$. The derived ages range from very young ages (7.4 Myr) to 437 Myr with a median age of 25 Myr. The color excess $E(B-V)$ are between $0.1-0.4$ mag. Star formation rates are $55-209$ $M_{\odot}$yr$^{-1}$. A comparison of the stellar populations is made between three LAEs and 88 LBGs selected at the same redshift, in the same observed field, and down to the same limit of the rest-frame UV luminosity. These three LAEs are the brightest and reddest samples among the whole LAE samples at $z=4.86$. The LAEs distribute at the relatively faint part of UV-luminosity distribution of LBGs. Deriving the stellar properties of the LBGs by fitting their SEDs with the same model ensures that model difference does not affect the comparison. It is found that the stellar properties of the LAEs lie on distributions of those of LBGs. On average, the LAEs show less dust extinction, and lower star formation rates than LBGs, while the stellar mass of LAEs nearly lies in the middle part of the  mass distribution of LBGs. However, the stellar properties of LAEs and LBGs are similar at the fixed UV or optical luminosity. We also examine the relations between the output properties from the SED fitting and the rest-frame \lya~equivalent width, but cannot find any significant correlation. 
\end{abstract}


\altaffiltext{1}{Based on data collected at Subaru Telescope, which is operated by the National Astronomical Observatory of Japan.}
\altaffiltext{2}{Department of Astronomy, Kyoto University, Sakyo-ku, Kyoto 606-8502, Japan}
\altaffiltext{3}{Department of Astronomy, School of Science, University of Tokyo, Tokyo 113-0033, Japan}
\altaffiltext{4}{Observatories of the Carnegie Institution of Washington, 813 Santa Barbara St., Pasadena, CA 91101}
\altaffiltext{5}{Okayama Astrophysical Observatory, National Astronomical Observatory of Japan, Okayama 719-0232, Japan}
\altaffiltext{6}{Department of Astronomy and Physics, Saint Mary's University, 923 Robie Street, Halifax, Nova Scotia, B3H 3C3, Canada}

\section{Introduction}\label{sec:intro}
There are two popular techniques for isolating galaxies at high redshift ($z\gtrsim3$): broadband selection (i.e., Lyman Break technique) and narrowband selection. Lyman Break Galaxies or LBGs are selected by the Lyman Break technique by taking the advantage of the spectral discontinuity due to the neutral hydrogen attenuation in the intergalactic medium at the rest-frame wavelength shorter than 912/1216 \AA. Via this method, a large number of high redshift galaxies are studied \citep[e.g.,][]{steidel1996, steidel1999, steidel2003, giavalisco1996, giavalisco1998, lowenthal1997, pettini1998, shapley2001, shapley2003, iwata2003, iwata2007, papovich2004, bouwens2004, bouwens2008, reddy2005, reddy2006, reddy2008, sawicki2005, sawicki2006, yoshida2006}. The other method is to select galaxies with strong \lya~emission lines which fall into a narrowband filter. This method is useful for selecting high-redshift galaxies that have a strong \lya~emission line. These objects are called Lyman Alpha Emitters (hereafter LAEs). \citet{partridge1967} proposed that primordial galaxies in the early stage of their formation should show a strong \lya~emission line. LAEs are thus expected to be young galaxies with low metallicity. Many surveys have been made to seach for galaxies with a strong \lya~emission at various redshifts ranging from 2.1 to 6.6 or even more \citep[e.g.,][]{hu1996, cowiehu1998, hu1998, hu2004, rhoads2000, rhoads2004, kudritzki2000, rhoads2001, malhotra2002, fynbo2001, ouchi2003, ouchi2008, fujita2003, shimasaku2003, shimasaku2004, shimasaku2006, kodaira2003, ajiki2003, ajiki2004, ajiki2006, taniguchi2005, venemans2002, venemans2004, iye2006, ota2007, nilsson2009, guaita2009}. 

In order to examine the evolutionary stage of LAEs and to reveal what kind of galaxy an LAE is, various properties of LAEs have been studied such as luminosity functions \citep[e.g.,][]{ouchi2003, ouchi2008, malhotra2004, shimasaku2006, kashikawa2006, gronwall2007, ota2007} or clustering properties \citep[e.g.,][]{ouchi2003, shimasaku2003, shimasaku2004, shimasaku2006, kashikawa2006, murayama2007}. Revealing the stellar population of LAEs is one of crucial studies for understanding their physical nature. In order to do that, spectral energy distribution (SED) of a galaxy is compared with stellar population synthesis models produced by varying the ages, metallicities, amounts of dust extinctions, SFRs, etc. The stellar population of a galaxy can be constrained from the best fit model. This method is known as SED fitting method. Recent studies show that LAEs are in wide ranges of ages (1Myr -- 1Gyr) and stellar masses ($10^6-10^{10}\Msun$) \citep[e.g.,][]{gawiser2006, gawiser2007, lai2007, lai2008, nilsson2007, finkel2007, finkel2008, finkel2009, pirzkal2007}. Among these studies, stacking analysis shows that LAEs at $z\sim3$ are free from dust or show modest dust extinction (E(B-V) $\leq$ 0.03 mag) \citep[e.g.,][]{gawiser2006, gawiser2007, lai2008, ono2009}, whereas SED fitting of an individual LAE at higher redshifts suggests that some of LAEs show significant dust extinction 
\citep[e.g.,][]{lai2007, 
finkel2009}. 

It is important to investigate the connection between LAEs and other galaxy populations selected by different method, i.e., LBGs that are selected based on UV continuum. Since these two methods suffer from different biases, the selected galaxies partially overlap and the relationship between them is not clear. \citet{shapley2001} divided LBGs at $z\sim3$ into 2 subsamples according to their ages obtained by SED fitting: "old subsample" with ages about 1 Gyr and "young subsample" with ages less than 35 Myr. They found that the old subsample shows a strong \lya~emission line, while young subsample do not. In other words, they found that LBGs with \lya~emission line are older than those without \lya~emission line. Confirming this statement, the recent work with a larger sample at the same redshift by \citet{kornei2009} shows that objects with rest-frame \lya~equivalent width larger than 20 \AA~seem to be older, lower in star formation rate, and less dusty than those without \lya~emission line. \citet{pentericci2007}, in contrast, found that at $z\sim4$ LBGs with \lya~emission are less massive and younger than those with no line. By considering the results from the SED analysis mentioned above, there may be an evolution of difference between LAEs and LBGs with redshifts \citep[e.g.,][]{shimizu2010}. Although there are some SED studies at higher redshift ($z\gtrsim5$) \citep{pirzkal2007, lai2007}, the relationship between LAEs and LBGs is still unknown. In this paper, we study the stellar populations of LAEs at $z=4.86$ from their rest-frame UV-to-optical spectral energy distributions. Down to the same UV luminosity, the derived stellar properties of LAEs are compared to those of LBGs by \citet{yabe}. They selected LBGs at the same redshift range in the same field and they used the same fitting code and synthesis spectral model assumptions. This direct comparison between LAEs and LBGs is expected to reveal their difference and to know what kind of properties make a galaxy an LAE. 

Data sources and photometry are described in section \ref{sec:data}. Section \ref{sec:select} explains the selection criteria, and the LAE candidates. The observed SEDs are constructed for each LAE candidate and compared to the models in section \ref{sec:sedfit}. The fitting results are shown in Section \ref{sec:results}. Section \ref{sec:compareLBG} is for comparisons between LAEs and LBGs. The summary is given in section \ref{sec:conclusion}. Through out this paper, we use AB magnitude system \citep{okegunn1983} and adopt a cosmology with parameters of $\Omega_{\rm m}=0.3$, $\Omega_{\Lambda}=0.7$, and $H_0=70$ km s$^{-1}$ Mpc$^{-1}$.\\

\section{Data Sources and Photometry}\label{sec:data}
\subsection{Optical data}\label{subsec:subarudata}

\subsubsection{Observations and data reduction}\label{subsub:reducesubaru}

\begin{figure}[t]
  \centering
  \includegraphics[trim =0mm 23mm 0mm 18mm, clip,angle = -90, width=7cm]{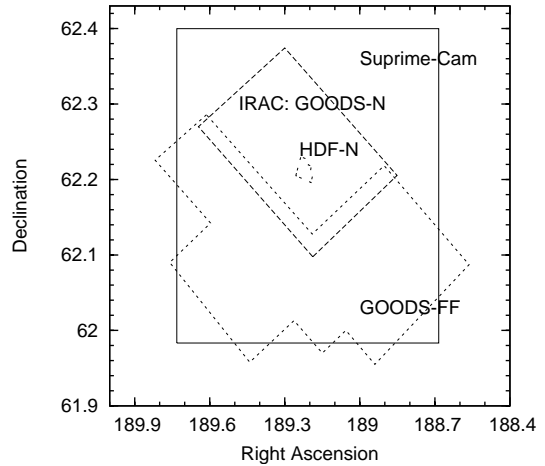}
  \caption[The field of observations both by Suprime-Cam and IRAC]
  {Observed fields. The solid line indicates the effective field observed with Suprime-Cam, whereas the long and short dashed lines show the GOODS-N and GOODS-FF fields observed with IRAC. Hubble Deep Field North (HDF-N) is shown as a reference at the center of the figure. 
  }
  \label{field}
\end{figure}

\begin{figure}
  \centering
  \includegraphics[trim=0mm 10mm 0mm 10mm, clip, angle=-90, width=9cm]{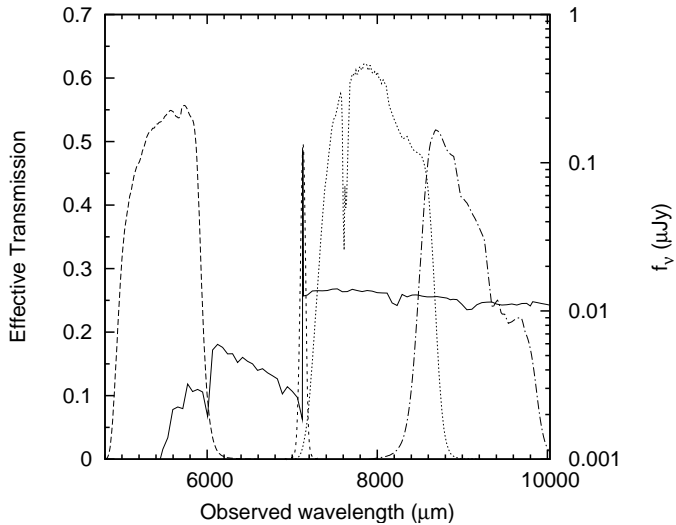}
  \caption[An example of LAE spectra and filter transmission curves]
  {Transmission curves of $V$, NB711, $I_c$, and \zmag~ bandpasses. CCD response, transmission of PFC, mirror reflectivity, and airmass (sec $z$ = 1.2) are included. Also shown is a model spectrum in the unit of $\mu$Jy (solid line) of a star forming galaxy at $z=4.86$ including \lya~emission line with a rest-frame equivalent width of 10\AA. The model is constructed by Bruzual \& Charlot (2003) stellar synthesis code. A constant star formation history and Salpeter IMF are assumed at an age of 12 Myr and at star formation rate of 1 $M_{\odot}$yr$^{-1}$. The attenuation by  intergalactic medium is applied with the prescription by Madau (1995). 
          }
  \label{laemodel}
\end{figure}

Optical data were obtained with Suprime-cam \citep{miyazaki2002} attached on the Subaru telescope \citep{iye2004}. The observed field is toward the Hubble Deep Field-North (HDF-N; \citealp{williams1996}) [RA(2000)=$12^{\rm h}36^{\rm m}49.^{\rm s}4$, Dec(2000)=+62$^{\circ}12'58''$] and is illustrated by solid line in Figure \ref{field}. Pixel scale of the CCD was $0.\arcsec20$. We used $NB711$ [$\lambda_{eff}=7126\rm{\AA}$, $\rm{FWHM}=73\AA$] and three broadband filters: $V$, $I_c$, and \zmag~to select $z=4.86$ LAEs. The effective redshift interval calculated from FWHM of $NB711$ is $4.83\leq z \leq 4.89$. Transmission curves, which include the responses of CCD, prime focus corrector (PFC), mirror, and airmass, of all filters used in this work are shown in Figure \ref{laemodel}. 
The observation with the NB711 filter was made on March 16, 2005. Images were taken with dithering of $\sim80\arcsec$ and the exposure time of 1200 seconds for each frame. With 23 exposures, we covered $\sim750~ {\rm arcmin}^2$ field of view. The total integration time was 7.7 hours. The weather condition was not so excellent. Seeing size during the observation was averagely $1.\arcsec5$. 

Data reduction was made by using SDFRED version 1.4.1, the software developed for Suprime-Cam data reduction \citep{yagi2002, ouchi2004}. After the bias subtraction, flat fielding, and distortion correction, image mosaicking was made by identifying $30-170$ non-saturated stars common in several object frames. The frame alignment and correction for flux/count and full width at half maximum (FWHM) were made based on these star data. A resulting FWHM of point sources in the mosaicked image is $\sim1.\arcsec7$. 

Astrometry was made based on USNO-A2 catalog by fitting with the third-order polynomial coefficients to $\sim1,000$ stars identified in the mosaicked image. The positional error is about $0.2\arcsec$ rms over the image. The magnitude zero point was derived based on the imaging data of two spectrophotometric standard stars (G191-B2B and HZ44) taken during the same observing night. As an independent check for photometry in $NB711$ images, we derived $NB711$ magnitudes of stars by interpolating their $V$, $I_c$, and \zmag~magnitudes. The derived magnitude zero points are in agreement within $\sim0.18$ mag. A 3$\sigma$ limiting magnitude at $2.\arcsec5$ diameter aperture is 26.16 mag for the $NB711$ image. 

The observations and data reductions of broadband images are described in detail by \citet{iwata2007}. The images were taken with $V$, $I_c$, and \zmag~filters in February 2001. The typical seeing during the observations was $\sim1.\arcsec1$. In addition, \citet{iwata2007} also used the imaging data taken from February to April 2001 by the University of Hawaii (UH) group \citep{capak2004}, who used the same instrument and filters. The final effective survey area is 508.5 arcmin$^2$. The limiting magnitudes of $V$, $I_c$, and \zmag-band images are 28.1, 26.8, and 26.6 mag, respectively ($3\sigma$ at $1.\arcsec6$ diameter aperture). The broadband images were all degraded to a PSF size of $1.\arcsec7$ to match that of the $NB711$ image. The $3\sigma$ limiting magnitudes of final PSF matched images at $2.5\arcsec$ diameter aperture are 27.3, 26.1, and 25.9 mag for $V$, $I_c$, and \zmag~images, respectively. 

\subsubsection{Photometric catalog}\label{subsub:photosubaru}

The photometric catalogs of the optical images were made by using SExtractor version 2.5.2 (Bertin \& Arnouts 1996). The positions of objects were extracted from the $NB711$ image and the photometry was then made at the extracted positions in all optical images via dual-image mode. In order to use the dual-image mode, we first have to make a position registration between the broadband images and the $NB711$ image. Position mapping between the images was made by using $geomap$ in IRAF\footnote{IRAF is distributed by the National Optical Astronomy Observatories, which are operated by the Association of Universities for Research in Astronomy, Inc., under cooperative agreement with the National Science Foundation.} based on positions of the bright but non-saturated stars detected in the images. Then the registration of the images was performed by $gregister$. The accuracy in making a position registration is within 0.2 pixels or 0.\arcsec04 rms. After image registration, we subtracted {\tt GLOBAL} background from each registered image individually\footnote{Based on our simulation with artificial objects, the processes of subtracting background before smoothing the images and re-subtracting the {\tt LOCAL} background in SExtractor give the better photometry.} and homogenized the broadband images so that they all have the same seeing size as the $NB711$ image within 0.\arcsec01 accuracy by $gauss$. The dual-image mode was then performed by using the following parameters. The images were filtered with a default convolution kernel (default.conv). The {\tt LOCAL} background was estimated using $64\times64$ pixel background mesh with $3\times3$ median filtering. An object was detected with a minimum of 5 connected pixels above $2.0\sigma$ minimum threshold. These parameters are found to maximize the number of detected objects in the $NB711$ image and minimize the detected numbers of objects in the negative $NB711$ images. 

In SED fitting process, the total photometry of the objects is necessary. The total magnitudes were obtained by applying aperture corrections to $2.\arcsec5$ diameter aperture magnitudes. We examined the best value of the aperture size and chose $2.\arcsec5$, which maximizes the signal-to-noise ratio (S/N). The aperture corrections were determined from the Monte Carlo simulations where artificial objects with PSF shape corresponding to the seeing size of $1.\arcsec7$ were put into the original image and were then detected by using the same SExtractor parameters. In the simulations, our objects were implicitly assumed to be point sources since their apparent size in the image is comparable to that of PSF. The PSF was made by stacking images of the point spread function that shows the stellarity index larger than 0.98 and has no nearby objects. The aperture correction factors are $-0.30\pm0.01$ and $-0.33\pm0.01$ mag for $I_c$ and \zmag~images, respectively, where the errors are the PSF uncertainties. Errors of the total magnitudes are the combination of $1\sigma$ error in making photometry of simulated objects and the uncertainties of correction factors. Because we do not use $V$ and $NB711$ images in SED fitting process, we do not compute the aperture corrections for them. 

\subsection{Mid-infrared data}\label{subsec:iracdata}

Mid-infrared images are obtained from deep observations with Infrared Array Camera (IRAC) on Spitzer Space Telescope (SST). We used the publicly available mid-infrared data in the Great Observatories Origins Deep Survey North field (\citealt{dickinson2003}, hereafter GOODS-N) centered at 12$^{\rm h}$36$^{\rm m}$54.$^{\rm s}$87, +62$^{\circ}14^\prime19.\arcsec2$ (J2000) provided by Spitzer Space Telescope Legacy Science program\footnote{http$\colon$//ssc.spitzer.caltech.edu/legacy/goodshistory.html}. The field covers an area of approximately $10' \times 16'$ or $\sim160$ arcmin$^2$ as shown in Figure \ref{field}. In this work, we used the First Data Release (DR1) and Second Data Release (DR2) of IRAC data consisting of imaging data in 3.6, 4.5, 5.8, and 8.0 \micron~bandpasses. Pixel scale of all images after drizzled is $0.\arcsec60~\rm{pixel}^{-1}$. Mean FWHM of PSF in IRAC GOODS-N images is $\sim$1.\arcsec7. 3$\sigma$ limiting magnitudes of IRAC 3.6, 4.5, 5.8, and 8.0 \micron~images at $2.\arcsec4$ diameter aperture are 25.9, 25.8, 23.7, and 23.6 mag, respectively. 

In addition to the very deep GOODS-N data, data in the flanking field (hereafter GOODS-FF) of GOODS-N are obtained to cover a part of the Subaru observation area. The GOODS-FF observations were carried out with IRAC camera in December 2005 and June 2006 (the Spitzer GO program (GO-20218); PI = Ikuru Iwata). The observations were made by pointing at 5 different positions on the sky around the GOODS-N field. The total exposure time for each passband is 4,000 seconds. Basic Calibrated Data (BCD) processed by the pipeline of Spitzer Science Center (SSC) were used. The images were drizzled using MOPEX software provided by SSC and combined together to get an improved mosaic image covering the whole observed area of $\sim 300$ arcmin$^2$. The mosaic image has a final pixel scale of $0.\arcsec61~\rm{pixel}^{-1}$ and a PSF size (FWHM) of 1.\arcsec7. The $3\sigma$ magnitude limits at $2.''4$ diameter aperture are 25.0, 24.6, 22.1, and 22.3 for IRAC 3.6, 4.5, 5.8, and 8.0\micron~bands, respectively. Combining the infrared data obtained in GOODS-N and GOODS-FF fields, we have the effective area of $\sim$400 arcmin$^2$ covering $\sim$80\% of Subaru images as seen in Figure \ref{field}. 

The photometry of IRAC was made based on the positions of LAEs at $z=4.86$, which are selected from the criteria described in the next section. Since the dimensions and the pixel scale of the narrow-band image and IRAC images are different, we could not use dual-image mode in SExtractor to determine aperture photometry in IRAC images. $Phot$ provided by IRAF was used instead to make aperture photometry at $2.\arcsec4$ diameter, which are examined to be the best aperture size since it maximizes the S/N ratio. Because the positions in the $NB711$ image are used as the reference, we made a registration of IRAC images before making IRAC photometry of the objects. The position alignment was made globally for each IRAC image by using positions of point sources detected in both $NB711$ and IRAC images. The estimated errors in determining the position shifts are 0.\arcsec1 (0.17 pixels) for IRAC images in GOODS-N field and 0.\arcsec2 (0.33 pixels) for those in GOODS-FF. Since the error of 0.\arcsec2 may not be negligible, we re-made the photometry of objects at the positions separated from the corrected positions by 0.\arcsec2 and found the magnitude difference to be less than 0.2 mag. This magnitude difference gives less than 3\% difference in the stellar masses derived from SED fitting. 

Aperture corrections for IRAC images were determined by the following process. The artificial objects with IRAC PSF are put into the IRAC images. The PSF was made from images of objects identified to be a stellar object in Subaru images. For the GOODS-N field, the correction factors are $-0.62$ mag and $-0.72$ mag in 3.6\micron~and 4.5\micron~bands respectively. The correction uncertainties due to the uncertainties in PSFs are 5\% and 6\% in 3.6\micron~and 4.5\micron~bands respectively. For GOODS-FF images, they are respectively $-0.70$ mag and $-0.73$ mag with 6\% and 7\% uncertainties in 3.6\micron~and 4.5\micron~bands. Errors in IRAC photometry are $1\sigma$ standard deviations of sky background and the uncertainties of correction factors. Since 5.8\micron~and 8.0\micron~images show too low signal-to-noise (S/N) ratio to place a useful upper limit in SED fitting process, we do not use them in further analysis. 
 
\section{Sample selection}\label{sec:select}
\begin{figure}
\centering
\includegraphics[angle=-90, clip,trim=0mm 25mm 0mm 23mm, width=7cm]{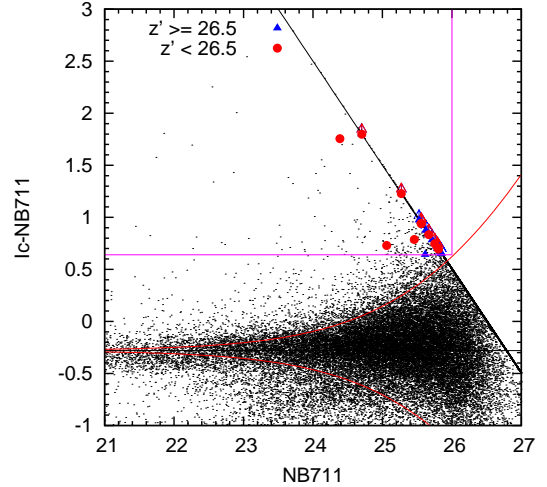}
\caption[Plot of narrow-band excess criteria]
{$I_c-NB711$ color versus $NB711$ $2.''5$ aperture magnitude. LAEs with $\zmag_{\rm{total}} < 26.5$ mag are shown with filled circles; filled triangles represent those with $\zmag_{\rm{total}} \geq 26.5$ mag. Arrows show the lower limits of $I_c-NB711$ taken at $2\sigma~I_c$ limiting magnitude. Note that we used $2\sigma~I_c$ limiting magnitude as a lower limit for objects whose $I_c$ magnitudes are fainter than that. A diagonal black line indicates $2\sigma$ limiting magnitude of the $I_c$ image. Red curves indicate the distributions of $3\sigma$ errors in measuring $I_c-NB711$ color for a source with a color of $I_c-NB711 = -0.28$ mag, the average color of all objects brighter than $NB711=26.0$ mag. The horizontal and vertical magenta lines show the $NB711-$excess criterion and $NB711=26.0$ mag limit, respectively. 
}
\label{nbrinb}
\end{figure}

29,675 objects were detected with $NB711 < 26.0$ mag ($3.5\sigma$ limiting magnitude). Figure \ref{nbrinb} shows $I_c-NB711$ color versus NB711 magnitudes of all objects (black dots). We selected $NB711-$excess objects from the following selection criteria:
\begin{eqnarray}
  I_c - NB711 &>& 0.64\label{crinbexcess}~\rm{and}\\
  I_c - NB711 &>& 3\sigma_{I_c - NB711},
\end{eqnarray}
where $3\sigma_{I_c-NB711}$ is a $3\sigma$ error in measuring $I_c-NB711$ color for a source with a color of $I_c-NB711 = -0.28$ mag, the average color of all objects with NB711 magnitudes brighter than 26.0 mag. Note that the magnitudes and colors of objects used in the sample selection are the values at 2.\arcsec5 diameter aperture, except for those noted otherwise. 
The first criterion corresponds to a rest-frame equivalent width (EW$_{\rm rest}$) of 10 \AA~ assuming a flat continuum ($f_{\nu} = \rm{constant}$). The EW cut of 10 \AA~used in this paper is lower than that commonly used in other LAE studies (20 \AA). Other studies \citep[e.g.,][]{ouchi2003, nilsson2009} compute the EW by using the average continuum flux density from the continuum both blueward and redward of the \lya~line, while the continuum in this work is extrapolated from the continuum redward of the \lya~line. Because the blueward continuum suffers from the attenuation by neutral hydrogen in the intergalactic medium (IGM) at high redshift, the continuum estimated by the latter way is expected to be higher than that in the usual case. Consequently, the EW computed in this paper is smaller than that from other studies, even though the object has the same  redward continuum. The \lya~EWs computed in the usual way are also shown in Table \ref{cons0.2}; most of the LAEs compared to LBGs show the EWs larger than 20\AA~when we adopt the commonly used way. 

By the above criteria, 667 objects were selected. Among them, there are low-redshift interlopers with strong emission lines such as \ha, H$\beta$, [OIII]$\lambda5007$, or [OII]$\lambda3727$. To rule them out, we first divided objects selected by the above criteria into two groups according to their brightness in \zmag~band. For candidates with $\zmag_{\rm{total}} < 26.5$ mag, we adopted the following criteria, which were used by Iwata et al. (2007; see their figure 2) to select LBGs at $z\sim5$: 
\begin{eqnarray}
\zmag_{\rm{total}} &<& 26.5,\\
V-I_c &>& 1.55 ~\rm{and}\\
V-I_c &>& 7.0(I_c-\zmag)+0.15,
\end{eqnarray}
where $\zmag_{\rm{total}}$ is the total magnitude in \zmag~band and colors are derived from 2.\arcsec5--diameter aperture magnitudes. These criteria have been confirmed by spectroscopic follow-up observations \citep{ando2004, ando2007, kajino2009}. For objects with $\zmag_{\rm{total}} \geq 26.5$ mag, we required non-detection in $V-$band magnitude ($V > 1\sigma$ limiting magnitude). As a result, 24 objects meet all criteria simultaneously down to $NB711 = 26.0$ mag as illustrated in Figure \ref{nbrinb}. Eleven of them show \zmag~magnitudes brighter than 26.5 mag. Note that $\zmag_{\rm{total}}$ here is slightly different from $\zmag_{\rm{magauto}}$ by \citet{iwata2007} (up to $\pm0.3$ mag at $\zmag=26.5$ mag) because the broadband images used in this work were smoothed to match the NB711 seeing size before making the photometry. 
\begin{figure}
\centering
\includegraphics[clip,width=8cm]{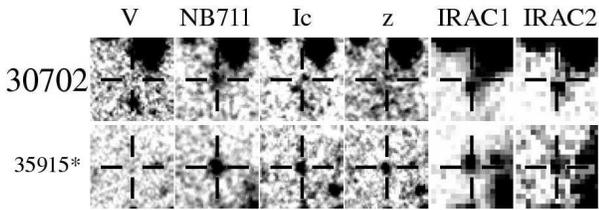}
\caption[Montage of class I LAEs]
{Montage images of group I LAEs. 
Size of each image is $10\arcsec\times10\arcsec$ with an LAE at the center of the image. North is at the top; east is to the left. IDs of objects are shown on the left of the figure. 
Note that the Subaru broadband images displayed here ($V$, $I_c$, and \zmag) are those before smoothing. An asterisk indicates the LAE with the spectroscopic redshift.}
\label{group1}
\end{figure}
\begin{figure} 
\centering
\includegraphics[clip,width=8cm]{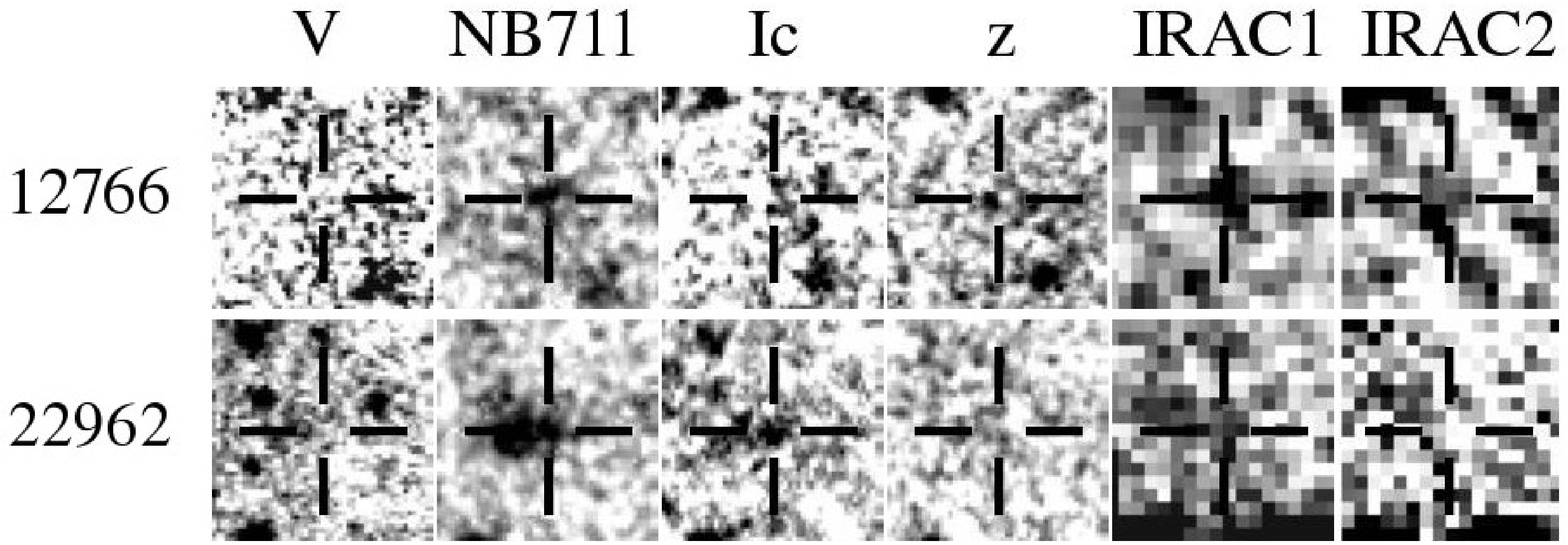}
\caption[Montage of class II LAEs]
{Same as Figure \ref{group1} but for group II LAEs.
}
\label{group2}
\end{figure} 
\begin{figure}
\centering
\includegraphics[clip,width=8cm]{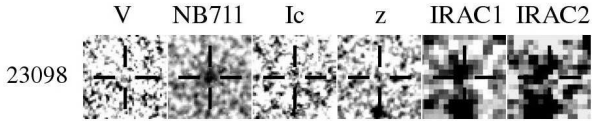}
\caption[Montage of class II LAEs]
{Same as Figure \ref{group1} but for group III LAE.
}
\label{group3}
\end{figure} 

\begin{figure}
\centering
\includegraphics[clip,width=8cm]{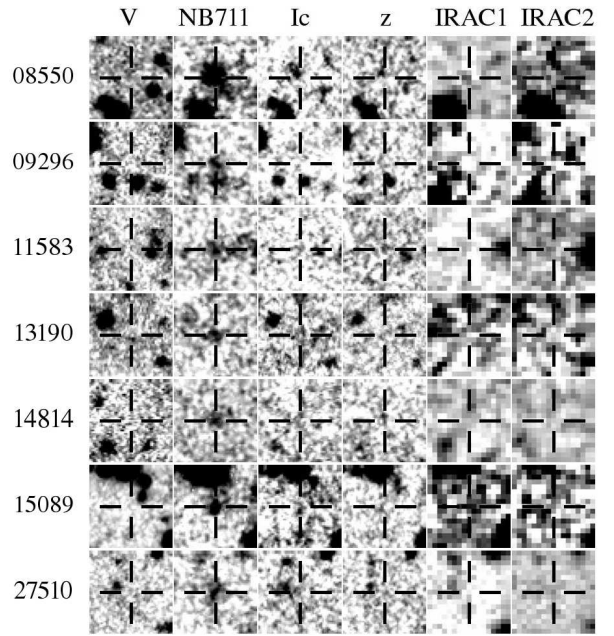}
\caption[Montage of class II LAEs]
{Same as Figure \ref{group1} but for group IV LAEs.
}
\label{group4}
\end{figure}

In order to study the properties of LAEs by SED fitting, the rest-frame optical photometry (i.e., IRAC photometry) is needed. 16 out of 24 LAEs are in IRAC images (hereafter IRAC sample). Since the angular resolution of IRAC images is not excellent, some LAEs are contaminated by nearby objects. In order to examine whether or not neighboring objects affect photometry of LAEs, we simultaneously fitted all neighboring objects within 10\arcsec~radius with GALFIT \citep{peng2002} and subtracted them from the original image. In the GALFIT process, we masked the LAEs out so that the accurate sky estimation can be obtained. Then we made aperture photometry of the LAEs from the residual images. We only selected LAEs of which magnitude difference between the original images and the residual images are less than 0.1 mag as isolated LAEs. Finally, we have 12 isolated objects. We divided these 12 LAEs into four groups based on their brightness in rest-frame UV and optical wavelengths. Group I is for two LAEs which are detected above $2\sigma$ limiting magnitudes in all bands used in SED fitting process (i.e., $I_c$, \zmag, IRAC 3.6\micron~ and 4.5\micron). One of them, \#35915, is spectroscopically confirmed to be at $z=4.82$. The emission line does not show a doublet profile in $\rm{R}\sim2000$ spectrum obtained by GMOS on Gemini telescope (Ouchi et al. in prep). With the given resolution, this can rule out the possibility that the object is an [OII] emitter. Moreover, there seems to be a continuum break at the wavelength shortward of the line. Two LAEs are classified as group II which is an LAE detected in two bands: one in Subaru optical and one in IRAC bands. One LAE is classified as group III which is not detected in $I_c$ and \zmag~bands but detected in both 3.6\micron~and 4.5\micron~bands. Group IV is for 7 LAEs that are not detected above $2\sigma$ limiting magnitudes in three or more bands. In SED fitting process, it is necessary that an LAE should be detected above $2\sigma$ magnitude limit in more than two bands; therefore, we can use only group I--III LAEs. Hereafter we call group I--III LAEs SED fitting sample. The montage images of LAEs are shown in Figure \ref{group1}, \ref{group2}, \ref{group3}, and \ref{group4} for group I, II, III and IV, respectively.\footnote{The adjacent object of \#22962, which is very bright in $NB711$ band, is a separate object and is likely to be a low-redshift emitter, because it is detected in $V$ band and does not satisfy the $V-I_c > 1.55$ criterion.} Aperture photometry of the isolated LAEs is summarized in Table \ref{phototable}. Neighboring objects are seen around some of group I-II LAEs, but note that the neighboring objects do not affect the photometry more than 0.1 mag. It is noteworthy that the object \#23098 in group III is undetected in all optical bands except for the $NB711$ image, while a bright object is seen close to \#23098 with a slight offset in IRAC 3.6\micron~and 4.5\micron~images. Our aperture photometry includes the light from this object. However, we cannot rule out the possibility that the object is a foreground red object. When we subtract the object, the aperture magnitudes in IRAC bands are fainter than $2\sigma$ limiting magnitudes. Thus it should be kept in mind that the SED of this object may be contaminated.

\begin{deluxetable*}{c c cccc cc}
\tablewidth{0pt}
\tabletypesize{\footnotesize}
\tablecaption{Aperture photometry of isolated LAEs\label{phototable}}
\tablehead{
\colhead{ID} & \colhead{Field} & \colhead{$V$\tablenotemark{a,b}} & \colhead{NB711\tablenotemark{a}} & \colhead{$I_c$\tablenotemark{a,c}} & \colhead{\zmag\tablenotemark{a,c}} & \colhead{3.6\micron\tablenotemark{a,d}} & \colhead{4.5\micron\tablenotemark{a,d}}\\
\colhead{} & \colhead{} & \colhead{(mag)} & \colhead{(mag)} & \colhead{(mag)} & \colhead{(mag)} & \colhead{(mag)} & \colhead{(mag)}
}
\startdata
\multicolumn{7}{l}{Group I}&\\
\hline
30702 & GOODS-N   & $28.37\pm0.69$ & $25.46\pm0.18$ & $26.25\pm0.36$ & $26.12\pm0.36$ & $24.97\pm 0.14$ & $25.36\pm 0.22$\\
35915 & GOODS-N   & $>28.51$            & $25.06\pm0.12$ & $25.79\pm0.25$ & $25.61\pm0.24$ & $25.06\pm 0.16$ & $25.60\pm 0.27$\\
\hline
\multicolumn{7}{l}{Group II}&\\
\hline
12766 & GOODS-FF & $>28.51$ & $25.81\pm0.23$ & $>26.50$            & $26.33\pm0.42$ & $25.12\pm 0.35$ & $>25.05$\\
22962 & GOODS-N  & $>28.51$ & $25.62\pm0.20$ & $26.26\pm0.37$ & $>26.38$            & $26.06\pm 0.35$ & $>26.24$\\
\hline
\multicolumn{7}{l}{Group III}&\\
\hline
23098 & GOODS-FF & $>28.51$ & $25.85\pm0.24$ & $>26.50$ & $>26.38$ & $24.42\pm 0.20$ & $24.21\pm 0.23$\\
\hline
\multicolumn{6}{l}{Group IV}&\\
\hline
08550 & GOODS-FF & $>28.51$ & $24.39\pm0.07$ & $26.14\pm0.33$ & $>26.38$            & $>25.42$ & $>25.05$\\
09296 & GOODS-FF & $>28.51$ & $25.83\pm0.24$ & $>26.50$            & $>26.38$            & $>25.42$ & $>25.05$\\
11583 & GOODS-FF & $>28.51$ & $25.57\pm0.19$ & $>26.50$            & $>26.38$            & $>25.42$ & $>25.05$\\
13190 & GOODS-FF & $>28.51$ & $25.78\pm0.23$ & $>26.50$            & $>26.38$            & $>25.42$ & $>25.05$\\
14814 & GOODS-FF & $>28.51$ & $25.84\pm0.24$ & $>26.50$            & $>26.38$            & $>25.42$ & $>25.05$\\
15089 & GOODS-FF & $>28.51$ & $25.27\pm0.15$ & $>26.50$            & $26.30\pm0.42$ & $>25.42$ & $>25.05$\\
27510 & GOODS-N   & $>28.51$ & $25.56\pm0.19$ & $>26.50$            & $>26.38$            & $>25.42$ & $>25.05$\\
\enddata
\tablenotetext{a}{Errors are $1\sigma$ values.}
\tablenotetext{b}{Upper limits are $1\sigma$ values at 2.\arcsec5 diameter aperture.}
\tablenotetext{c}{Upper limits are $2\sigma$ values at 2.\arcsec5 diameter aperture.}
\tablenotetext{d}{Upper limits are $2\sigma$ values at 2.\arcsec4 diameter aperture.}
\end{deluxetable*}

\section{Selection bias}\label{sec:bias}
\begin{figure*}
\begin{center}
\begin{tabular}{ccc}
\includegraphics[angle=-90, width=5cm]{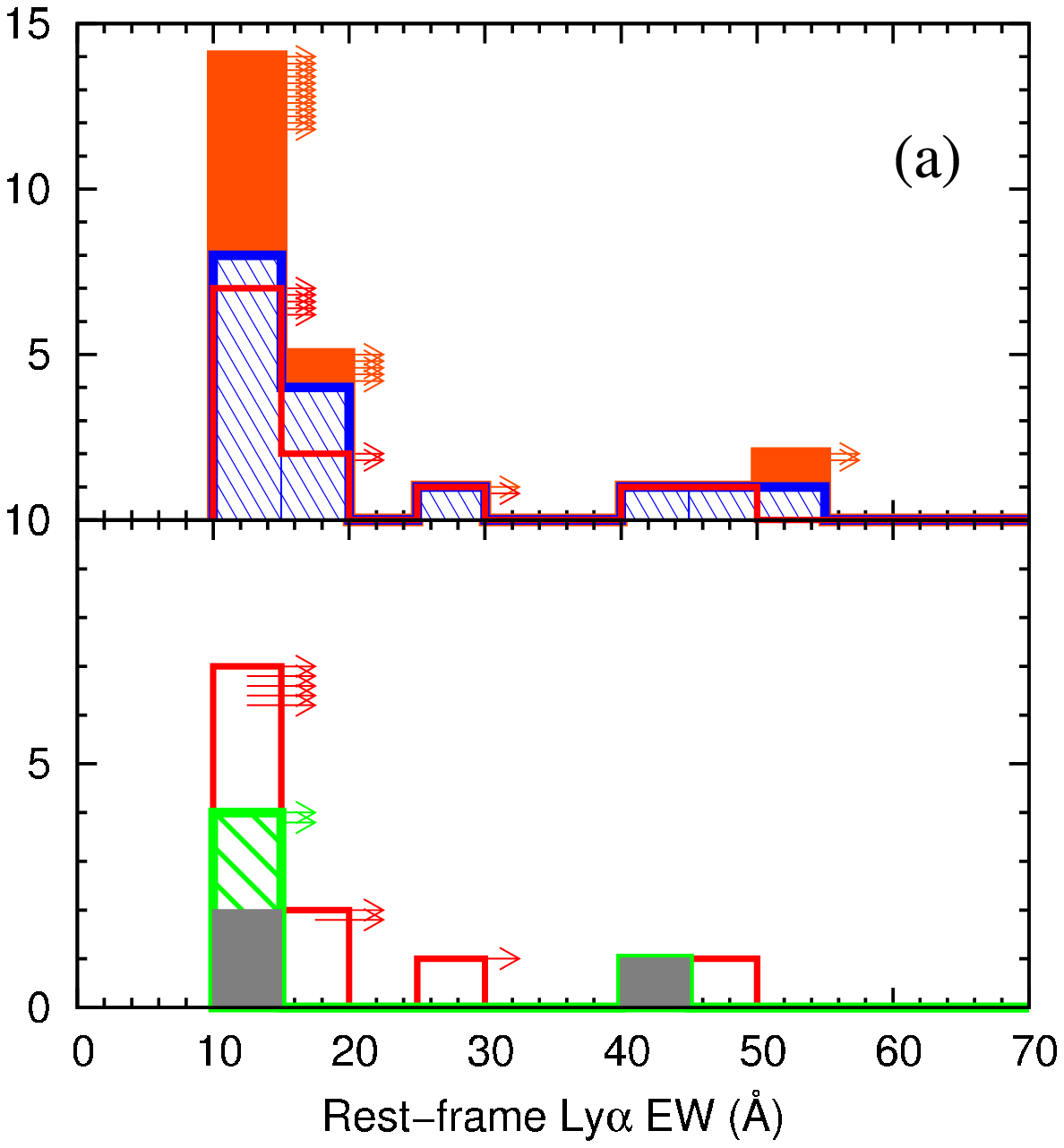} &
\includegraphics[trim=66mm 40mm 0mm 0mm, angle=-90, width=5cm]{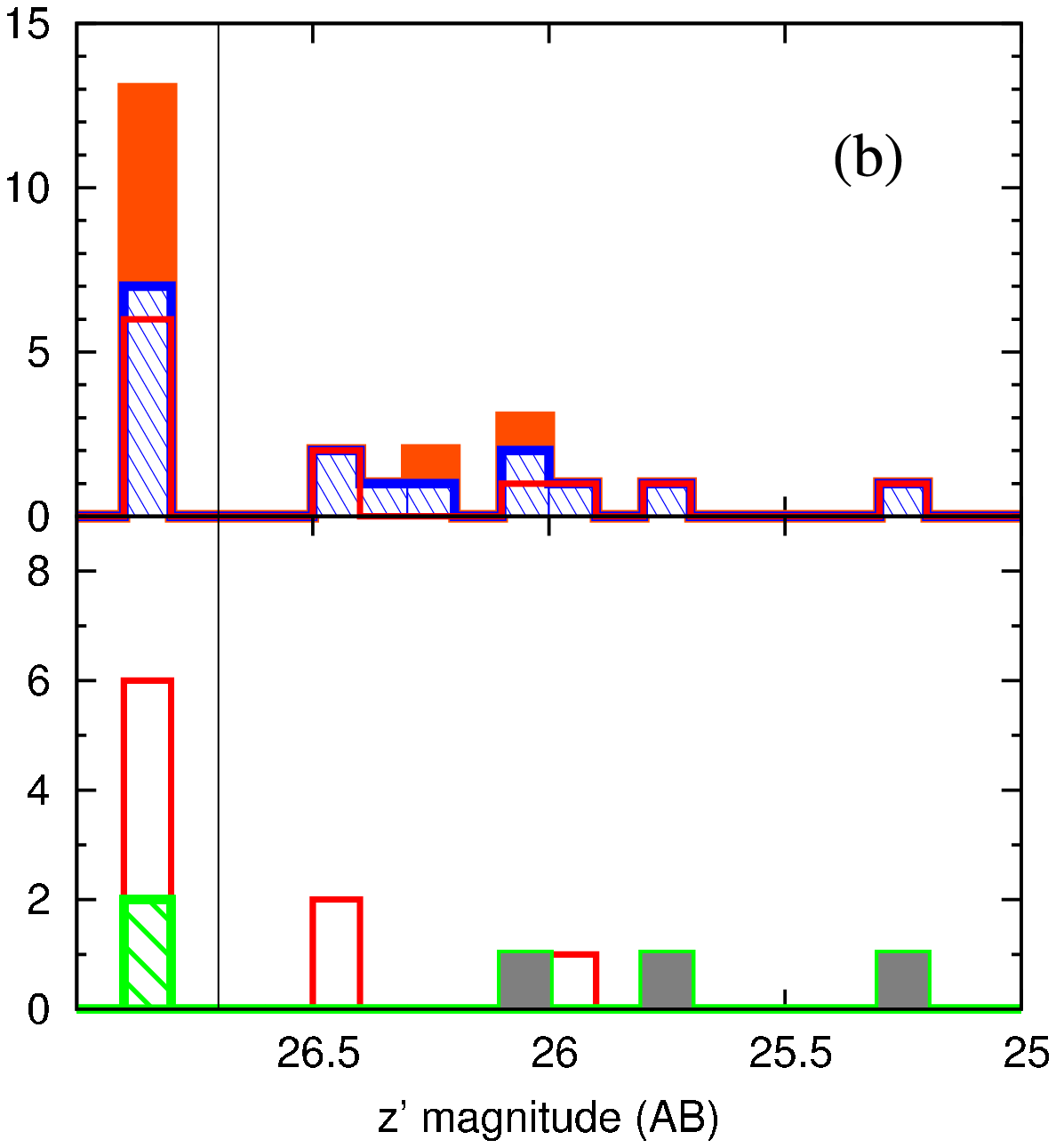} &
\includegraphics[angle=-90, width=5cm]{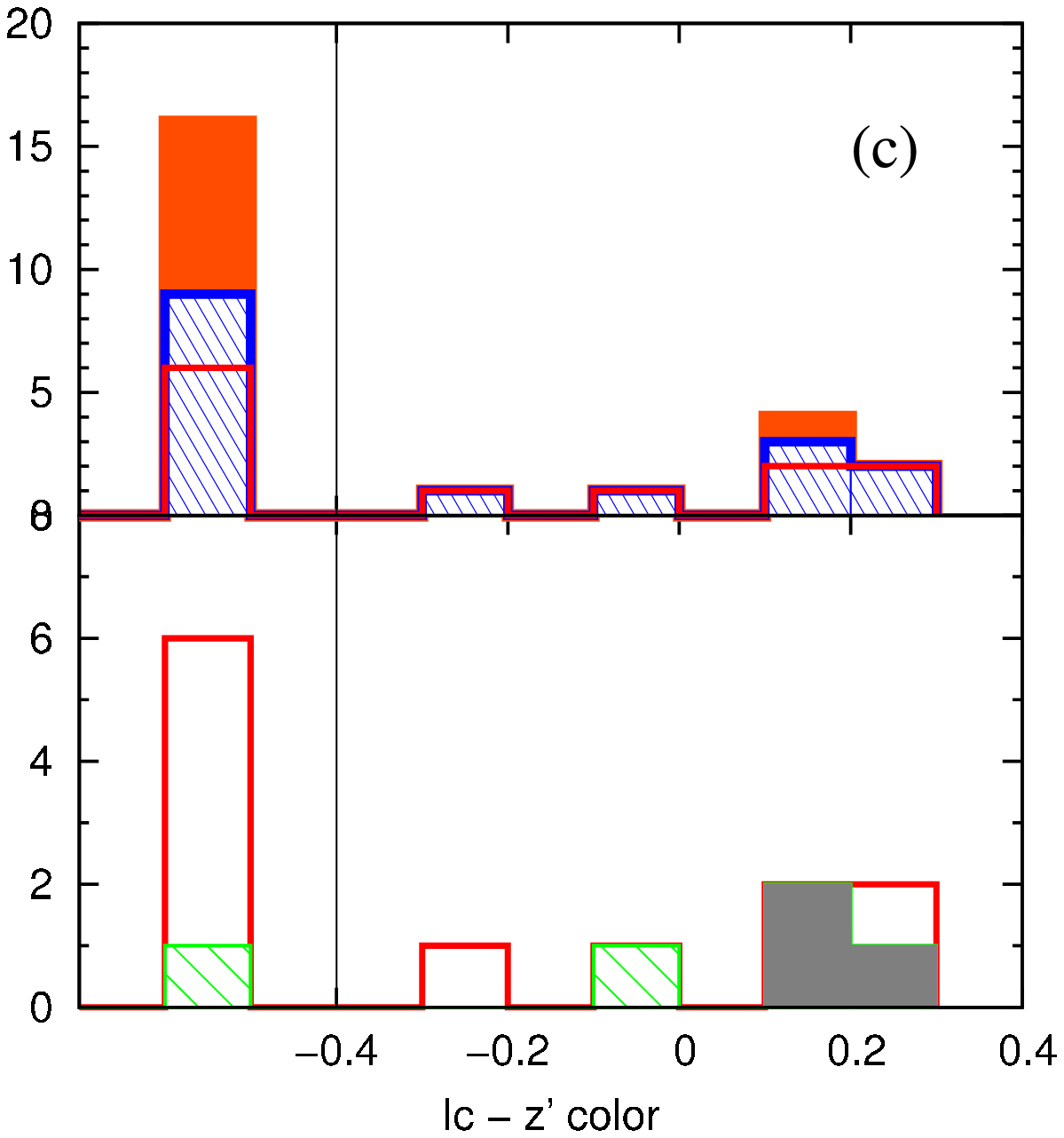} \\
\end{tabular}
\caption{Distributions of rest-frame \lya~EW, \zmag--band magnitude, and $I_c-\zmag$ color of LAE samples. We divide each figure into 2 panels for display purpose. In the top panel, an orange histogram shows the distribution of all 24 LAEs at $z=4.86$, while the blue and red histograms are for the IRAC samples and the isolated LAEs (group I--IV), respectively. In the bottom panel, red histograms refer to the isolated LAEs (the same as in the top panel). Group I--III LAEs are shown in green histograms. Distributions of group I LAEs and \#12766 from group II LAEs (the compared LAE sample; see section \ref{sec:compareLBG}) are shown in grey histograms. Arrows shown in panel (a) represent the lower limits in \lya~EWs of LAEs; the number of arrows directly corresponds to the number of LAEs with \lya~EW lower limits. Note that the leftmost bins in panels (b) and (c) indicate, respectively, the LAEs fainter than $\zmag = 26.5$ mag and LAEs that are not detected above $2\sigma$ limiting magnitudes in neither $I_c$ nor \zmag~bands.}
\label{bias}
\end{center}
\end{figure*}

In this section, we check the sample bias in two points of views: bias on selecting the isolated LAEs and bias on choosing the SED fitting sample. Figure \ref{bias}(a) shows the distributions of rest-frame \lya~equivalent widths of all LAEs, IRAC sample, isolated LAEs (group I--IV), SED fitting sample (group I--III), and the compared LAE sample (see section \ref{sec:compareLBG}). The rest-frame \lya~EWs are calculated by assuming that all LAEs are at $z=4.86$ and the \lya~emission line falls into the center of $NB711$ band (Table \ref{cons0.2}). Note that the values may be underestimated if the LAEs are not exactly at $z=4.86$, i.e., the \lya~emission line does not fall into the center of $NB711$ band. In addition, the EWs of 20 objects (of all 24 LAEs) are more underestimated due to the $2\sigma$ upper limit on $I_c$ magnitudes. For the spectroscopically confirmed LAE (\#35915), the \lya~emission line is in the $NB711$ band, but is out of $NB711$ FWHM range. By considering the transmission curve of the $NB711$ filter, the corrected EW is estimated to be 40 \AA~(Table \ref{cons0.2}) and is used hereafter. As a cross-check, we also examine the EW from the spectrum and find that it is consistent with that obtained from the imaging. The EWs of LAEs distribute from 10 \AA~(the selection limit) to 55 \AA. Figures \ref{bias}(b) and (c) show the distributions of \zmag~magnitudes and $I_c-\zmag$ colors, respectively. Leftmost bins in Figures \ref{bias} (b) and (c) represent LAEs with \zmag~magnitudes fainter than 26.5 mag and with magnitudes fainter than $2\sigma$ limiting magnitudes in both $I_c$ and \zmag~bands, respectively. The large number of LAEs in the leftmost bins of both figures imply that most of LAEs have faint UV continuum. As seen in Figures \ref{bias} (a)--(c), the isolated LAEs (red open histogram) occupy the same range of distributions as all sample at $z=4.86$ (orange histogram). In order to check whether or not selecting isolated sources in IRAC images can cause any bias on properties of the whole LAEs, we applied the Kolmogorov--Smirnov test (KS test) to the distributions of properties of all LAEs at $z=4.86$ and the isolated LAEs. We could not reject the null hypothesis that the rest-frame \lya~EW, \zmag~magnitude, and $I_c-\zmag$ color distributions of these two samples are drawn from the same populations at more than 95\% confidence levels. However, we could not use all isolated sample in SED fitting because of their faintness either in rest-frame UV or optical wavelengths. To investigate if SED fitting in the following section can represent all populations of LAEs at this redshift, we have to compare the SED fitting sample (group I--III, green histogram) to all LAEs (orange histogram). It is seen in Figure\ref{bias} (a) that group I--III LAEs represent both high-- and low--EW LAEs. However, we can see from Figures \ref{bias} (b) and (c) that the SED fitting sample are biased toward the bright UV luminosity and red $I_c-\zmag$ colors as we neglect the leftmost bins where magnitudes are unreliable. In addition, Table \ref{phototable} shows that SED fitting sample is relatively bright in IRAC bands as compared to all isolated LAEs implying the brighter luminosity in rest-frame optical wavelength. Thus we conclude that the SED fitting and the results hereafter are for the LAEs which have the brighter rest-frame UV and optical magnitudes, and the relatively redder UV colors. 

\section{Stellar population synthesis model and SED fitting}\label{sec:sedfit}

In this paper, we intend to compare stellar populations of the LAEs to those of LBGs at the same redshift by \cite{yabe}. In order to make a fair comparison, we used the same stellar population synthesis model as that used in \citet{yabe}. The model SEDs were obtained by Bruzual $\&$ Charlot synthesis code (2003, hereafter BC03). We used Padova 1994 evolutionary track as recommended by BC03. Salpeter(1955) initial mass function (IMF) with lower and upper mass cutoffs of 0.1 and 100 \Msun is assumed. We made models by fixing the metallicity at 0.2\Zsun\footnote{Fixing the metallicity at the lower abundance (0.005\Zsun) does not significantly change the fitting results. The stellar masses derived from the lower metallicity model differ from those derived from 0.2\Zsun~model by $\sim10\%$ at most. The average differences of age, color excess, and star formation rate are $\sim50\%$, $\pm0.2$ mag, and $\sim5\%$, respectively.} and assuming the constant star formation history. BC03 uses quasi-logarithmic 221 time steps from 0.1 Myr to 20 Gyr. Time steps were adopted to 51 logarithmic steps both to reduce the calculation time and to avoid dealing with an unequally spaced scale of the original 221 models. The age of the Universe at $z\sim5$ is $\sim1.2$ Gyr. However, as a cross-check on the fits, we allowed the age up to the oldest one available in BC03. The effect of dust attenuation is taken into account by using Calzetti extinction law (Calzetti et al. 2000) by changing the color excess $E(B-V)$ from 0.0 mag to 0.8 mag with a 0.01 step. Attenuation by the IGM is calculated with the prescription by Madau (1995). The model spectra were then convolved with the appropriate filter transmission curves to give model fluxes. This is exactly the same as done in \citet{sawickiyee1998}. Except for one spectroscopically confirmed LAE (\#35915; $z=4.82$), the redshift was fixed at $z=4.86$ under assumption that \lya~ emission of our LAE sample is detected at the center of the NB711 bandpass. 

Some of the observed SEDs including one spectroscopically confirmed LAE show a significant excess in the 3.6\micron~band as compared with the magnitude in the 4.5\micron band. This is likely to be due to the \ha~ emission line, which falls into IRAC ch1 (3.6\micron) at this redshift. The excess is also seen in some fraction of $z\sim5$ LBGs \citep{yabe}. The SED model including \ha~emission shows a better fit \citep{chary2005, finkel2008, yabe}. As adopted by \cite{yabe}, the spectrum of \ha ~emission is included in the synthesis model spectrum by the following process. The luminosity of \ha ~emission is calculated from the star formation rate of the model by using the relation by Kennicutt (1998).\footnote{Adopting this relation may be disputable. Dependences on metallicities and differences of extinction to stellar continuum and nebular emission are discussed by Yabe et al. (2009; their Appendix B).} The dust extinction to the line (Calzetti et al., 2000) is assumed to be the same as that to stellar component.\footnotemark[6] The \ha~ flux density is finally put into each model SED. The existence of \ha~ emission line in the models makes the fit better without adding a free parameter \citep{yabe}. 

We use the SEDfit software (Sawicki, in prep.), which is an evolved version of the SED-fitting software used in \cite{sawickiyee1998} and subsequent papers, including the $z\sim5$ LBG study by \cite{yabe}.  
After we produce a suite of model fluxes by means described above, the best-fitting model is found by means of a maximum likelihood test for each object. For objects detected in all bands (i.e., our group I objects), this test is the standard $\chi ^2$ minimization.  For objects that are undetected in some of the bands (i.e., group II and III objects), SEDfit employs a modification of the $\chi^2$ formalism: as is standard in the $\chi^2$ approach, for the filters in which the object is detected the maximum likelihood calculation considers the likelihood that the detected data deviate from the model given the uncertainties; for the undetected filters, the calculation adds in the likelihood that a given model would/would not have been detected given the upper limit of the non-detection. For more details on this technique see Sawicki (in prep.).

In this paper, we constructed an observed SED of each individual object from the photometry in $I_c$, \zmag, IRAC 3.6\micron, and IRAC 4.5\micron~bands. As explained in section \ref{sec:data}, we do not use $V$-band photometry in the SED fitting process in order to avoid the uncertainty due to IGM absorption. IRAC 5.8\micron~and 8.0\micron~photometry is not used because of the low S/N ratio of the images. Including the upper limits of the photometry in IRAC 5.8\micron~and 8.0\micron~bands does not usefully improve the fitting. Free parameters in our fitting process are age, color excess, and scaling normalization. The SFR is obtained from the scaling normalization. We then find the stellar mass by using the age and SFR \citep{sawickiyee1998}. 

\section{Results}\label{sec:results}
\begin{deluxetable*}{c c c c ccc c c}
\tabletypesize{\footnotesize}
\tablewidth{0pt}
\tablecaption{The best fit results \label{cons0.2}}
\tablehead{
\colhead{ID} & \colhead{Field} & \colhead{rest \lya~EW\tablenotemark{a}} & \colhead{log[Mass]} & \colhead{log[Age]} & \colhead{E(B-V)} & \colhead{log[SFR]} & \colhead{$\chi^2_\nu$} & \colhead{$q$\tablenotemark{b}}\\
\colhead{} & \colhead{} & \colhead{(\AA)} & \colhead{(\Msun)} & \colhead{(yr)} & \colhead{(mag)} & \colhead{(\Msun yr$^{-1}$)} & \colhead{} & \colhead{}
}
\startdata
\multicolumn{7}{l} {Group I}&\\
\hline
30702 & GOODS-N  & 14 (51) & $9.61_{-0.37}^{+0.99}$ & $7.49_{-0.62}^{+2.50}$ & $0.28_{-0.28}^{+0.10}$ & $2.15_{-1.39}^{+0.33}$ & $0.02$ & $1.32$\\
35915 & GOODS-N  & 40 (202) & $9.39_{-0.37}^{+0.72}$ & $7.39_{-0.62}^{+1.87}$ & $0.20_{-0.20}^{+0.09}$ & $2.03_{-1.08}^{+0.32}$ & $0.08$ & $1.16$\\
\hline
\multicolumn{7}{l} {Group II} &\\
\hline
12766 & GOODS-FF & $>11 (>44)$ & $9.29_{-1.40}^{+1.59}$ & $6.97_{-1.35}^{+3.33}$ & $0.30_{-0.3}^{+0.14}$ & $2.32_{-1.66}^{+0.37}$ & $44.90$ & $<1.44$\\
22962 & GOODS-N  & $10 (39)$ & $8.60_{-0.71}^{+1.23}$ & $6.87_{-0.83}^{+2.50}$ & $0.14_{-0.14}^{+0.10}$ & $1.74_{-1.15}^{+0.28}$ & $46.63$ & $2.27$\\
\hline
\multicolumn{7}{l} {Group III}&\\
\hline
23098 & GOODS-FF & $>10 (>11)$ & $10.69_{-0.60}^{+1.01}$ & $8.64_{-1.56}^{+1.66}$ & $0.40_{-0.32}^{+0.58}$ & $2.14_{-1.06}^{+1.40}$ & $47.05$ & $<1.01$\\
\enddata
\tablenotetext{a}{The rest-frame \lya~EWs are determined by assuming that all LAEs are at $z=4.86$ except for \#35915. Values in parentheses indicate the EW in case of using broadband images both redward and blueward of \lya~wavelength to estimate the continuum. }
\tablenotetext{b}{The clumpiness parameter (see text for further details).}
\end{deluxetable*}

\begin{figure}
\centering
\includegraphics[angle=-90, width=8cm]{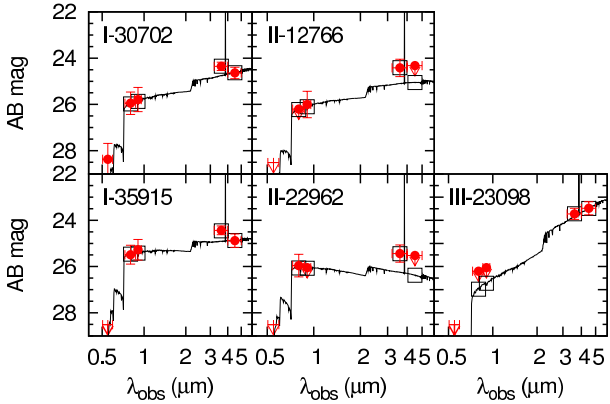}
\caption[Plot of observed SEDs and best fit model spectra]
{Observed SEDs and the best-fitting models for group I--III LAEs. In each panel, the observed SED is shown with filled circles; the best-fitting model SED is indicated by opened squares. Solid line represents the best-fitting model spectrum with \ha~emission line. Arrows indicate the 2$\sigma$ upper limits. V-band photometry is also shown in the figures but is not used in SED fitting.}
\label{sed1}
\end{figure}

The fitting results are summarized in Table \ref{cons0.2}. Errors in the table are at $68\%$ confidence level and are determined as follows. The Monte Carlo realizations for each object are performed; we vary the input fluxes within their photometric uncertainties, rederive the best-fitting model and repeat it 100 times. The error on an individual parameter is then 68\% range of the realizations in that parameter. The best-fitting model spectra are shown with the observed SEDs in Figure \ref{sed1}. 

\begin{figure*}[ht]
\centering
\begin{tabular}{cc}
\includegraphics[trim = 2mm 21mm 4mm 20mm, width = 6cm, angle = -90]{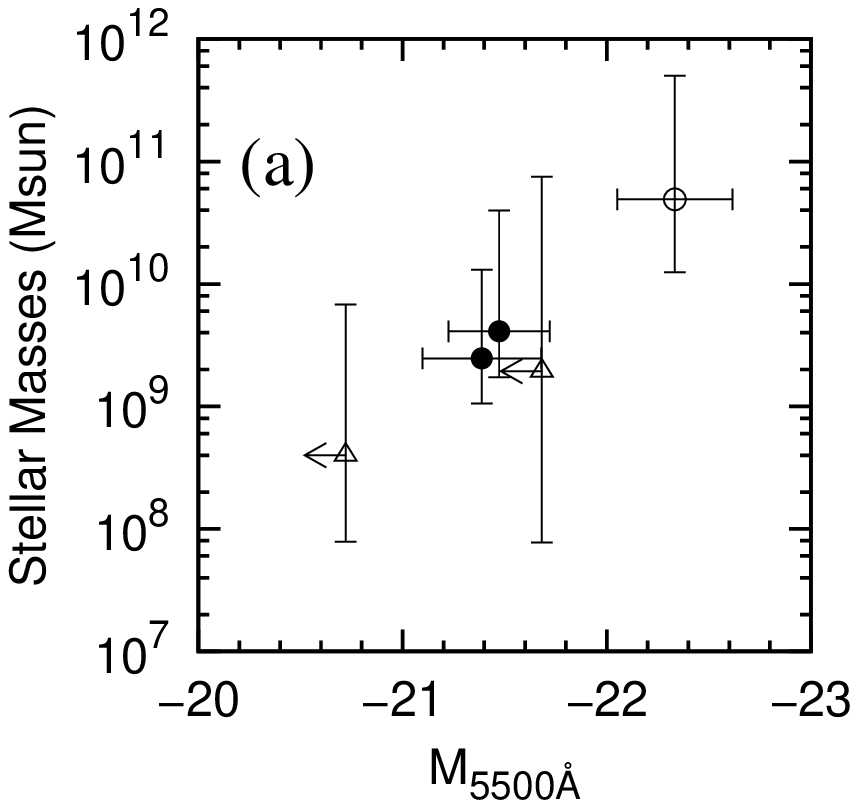} &
\includegraphics[trim = 2mm 21mm 4mm 20mm, width = 6cm, angle = -90]{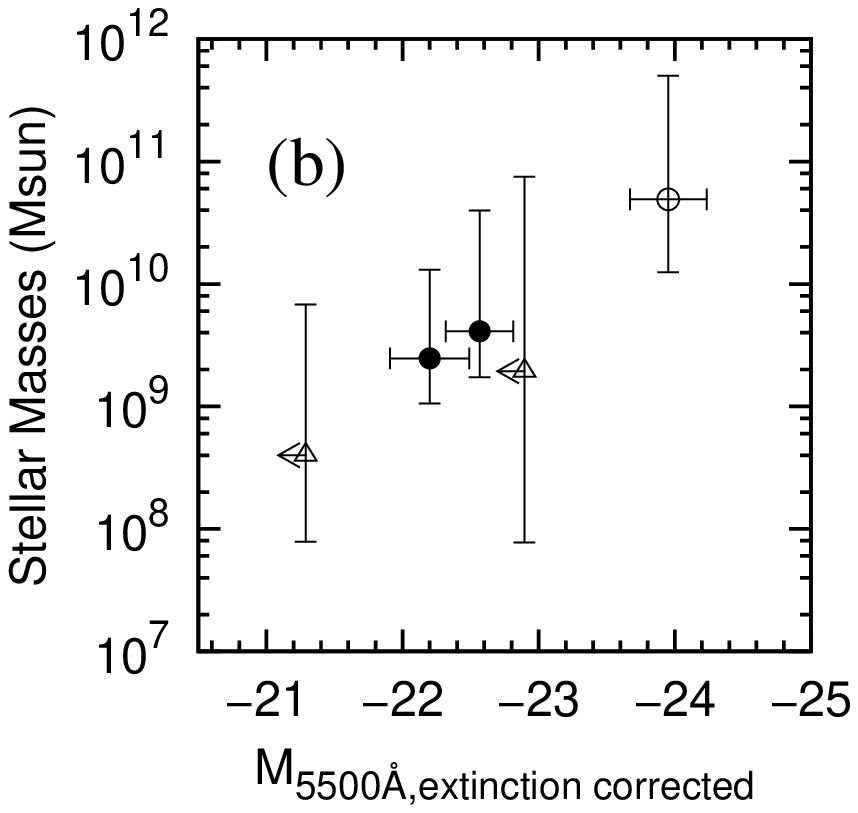} \\
\includegraphics[trim = 2mm 21mm 4mm 20mm, width = 6cm, angle = -90]{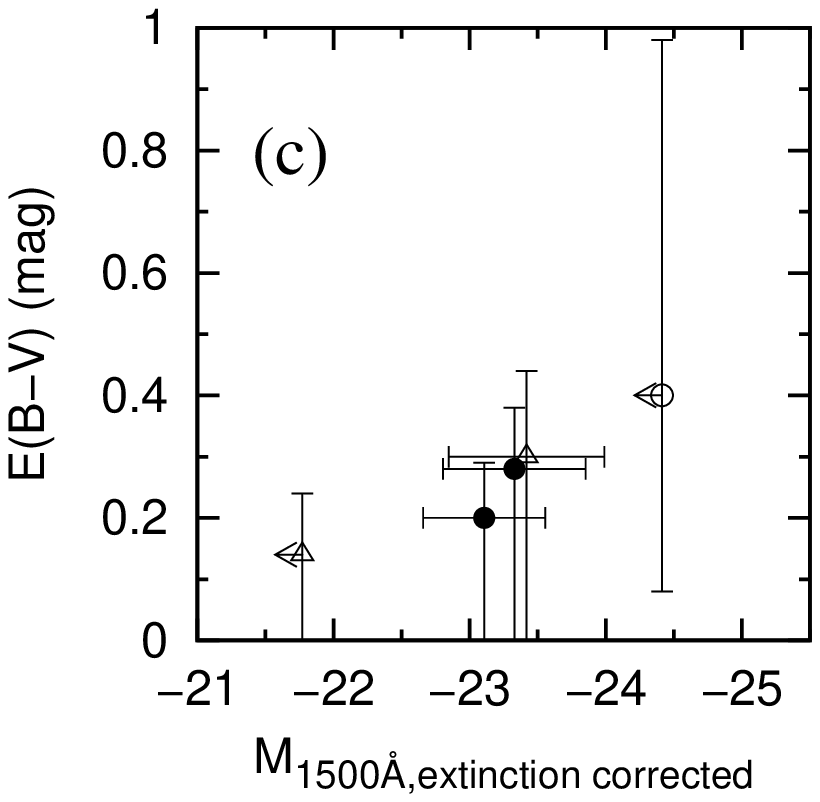} &
\includegraphics[trim = 2mm 21mm 4mm 20mm, width = 6cm, angle = -90]{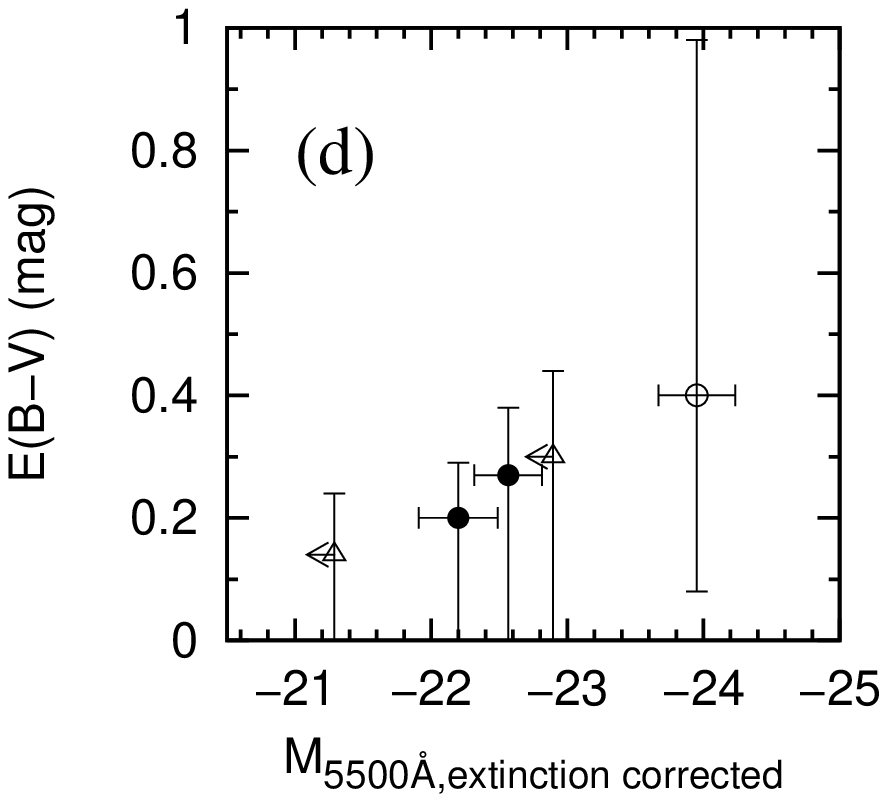} \\
\end{tabular}
\caption{Plots of derived stellar masses and color excesses of group I--III LAEs at $z=4.86$ against the rest-frame UV or optical absolute magnitudes. Filled circles, open triangles, and open circles represent group I, group II, and group III LAEs, respectively. Vertical error bars represent the 68\% errors as described in text. Arrows indicate the objects whose absolute magnitudes were calculated from the upper limits. }
\label{massopt}
\end{figure*}

The derived stellar masses are ranging from $10^8-10^{11}$ \Msun~with the median value of $2.5 \times10^9\Msun$. A typical error on the stellar mass is $\sim0.9$ dex. By considering only group I LAEs which are detected in all bands used in the SED fitting, the stellar mass is in the order of $10^9\Msun$ with the smaller typical error of $\sim0.6$ dex. A relationship between the stellar masses and the rest-frame optical absolute magnitudes is shown in Figure \ref{massopt}(a); the optical absolute magnitudes were calculated from the [4.5\micron] magnitudes by assuming $f_{\lambda}\propto \lambda^{\beta}$ with $\beta$ derived from the $\zmag-\rm{[4.5\micron]}$ color of an individual object. For LAEs with upper limit, we used the upper limit to estimate the magnitude. 
Despite the small size of the sample, the rest-frame optical luminosity is likely to be a tracer of the stellar mass. The relationship between the stellar mass and extinction-corrected optical absolute magnitude also shows a correlation (Figure \ref{massopt}(b)). A mass-to-light ratio is not perfectly linear; it becomes larger in the brighter LAEs. This trend is similar to the results of LBGs at the same redshift by \cite{yabe} which will be discussed in more details in the next section. From the relationship between the stellar mass and the rest-frame optical luminosity, it is implied that most LAEs at $z=4.86$, which could not be detected above $2\sigma$ limiting magnitudes in $4.5\micron$ band (Table \ref{phototable}), are likely to have stellar masses lower than $10^9\Msun$. 

\begin{figure*}[htbp]
\centering
\begin{tabular}{c}
\includegraphics[width = 5.5cm, angle = -90]{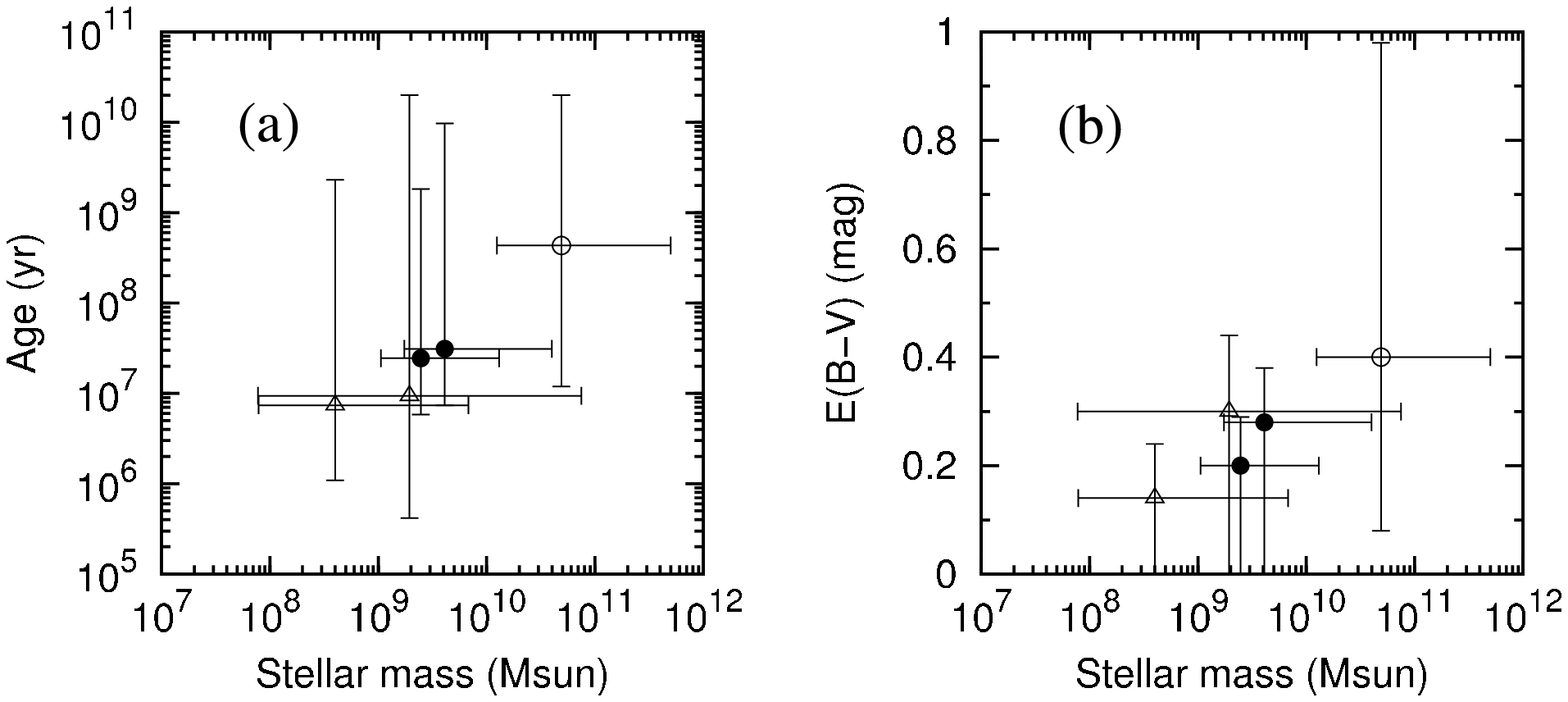} \\
\includegraphics[width = 5.5cm, angle = -90]{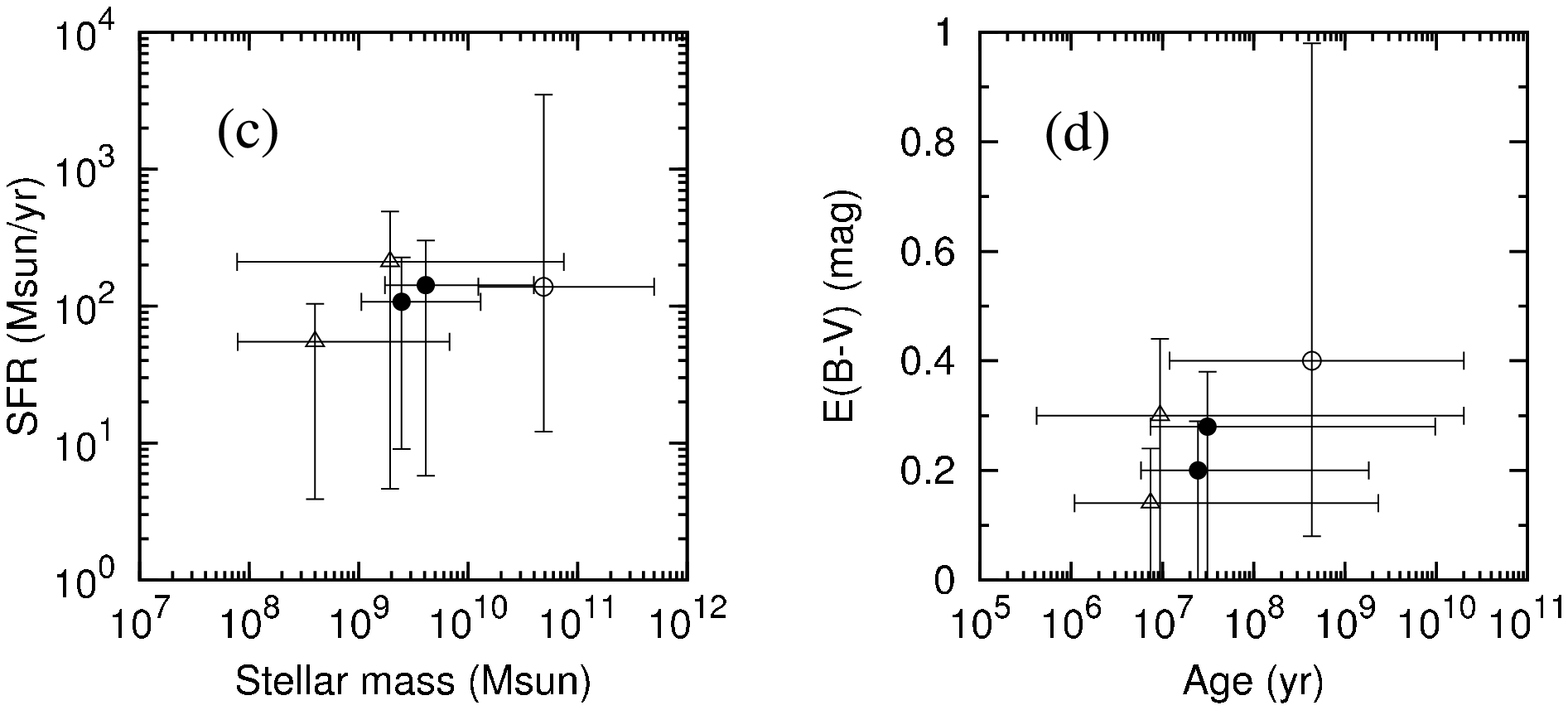} \\
\includegraphics[width = 5.5cm, angle = -90]{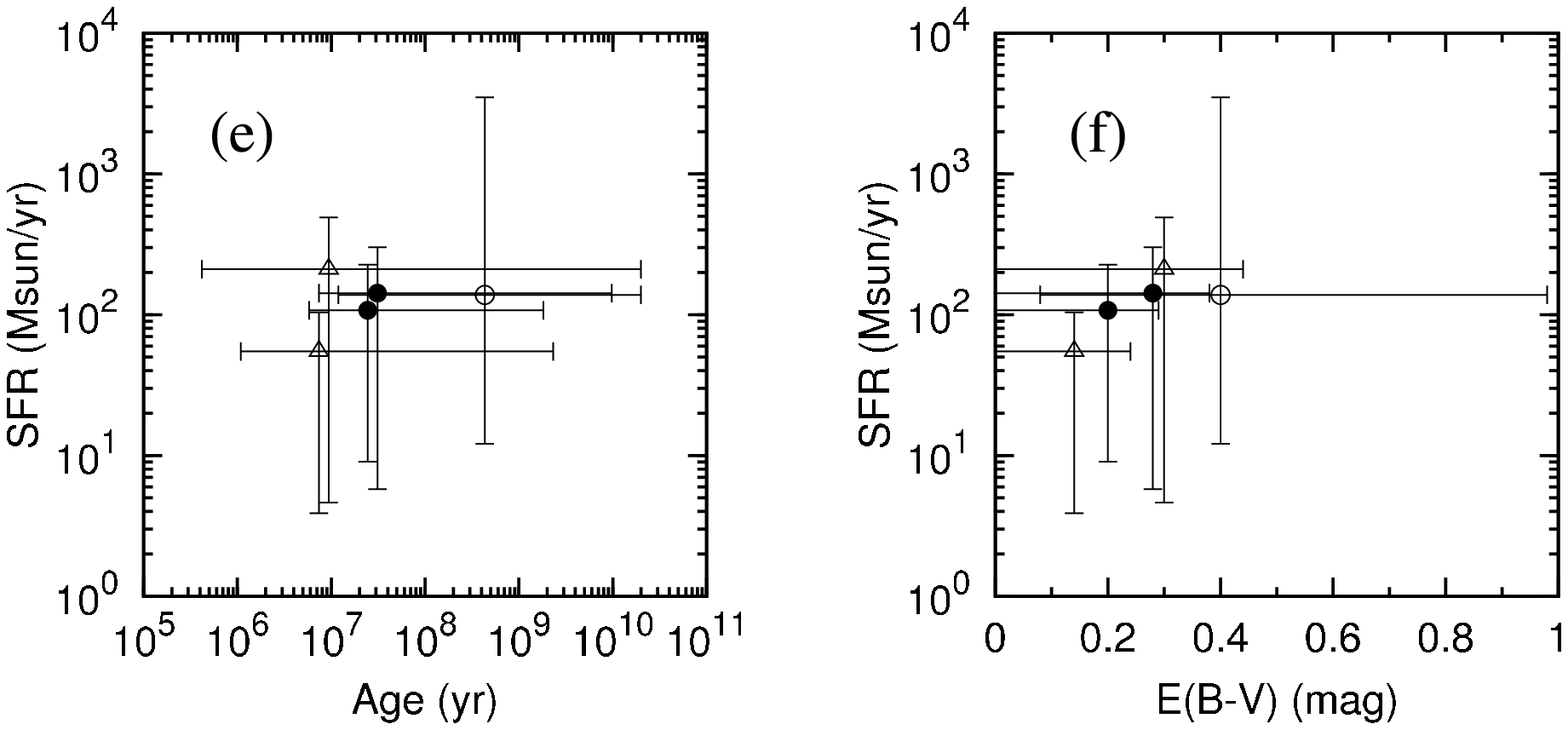} \\
\end{tabular}
\caption{Relations of output parameters from SED fitting for group I--III LAEs at $z=4.86$. Filled circles, open triangles, and open circles represent group I, II and III LAEs, respectively.}
\label{propmontage}
\end{figure*}

The best-fitting stellar ages are in a wide range from 7.4 Myr to 437 Myr with the median age of 25 Myr. Although we did not limit the model age at the age of the universe at $z=4.86$ ($\sim1.2$ Gyr), there is no LAE with the best-fitting age exceeding the cosmic age. The typical error on age of each object is very large ($\sim1.7$ dex). The uncertainties of ages cover most of the age range available in the models. Thus it is difficult to draw any conclusion on the correlation of them. 

The color excesses $E(B-V)$ range from 0.14 mag to the value as high as 0.40 mag. The median value is 0.27 mag. The typical error on $E(B-V)$ is $\sim0.2$ mag. It is seen in Figures \ref{massopt}(c) and (d) that dust extinction seems to show some correlation with both rest-frame UV and optical magnitudes. However, the trends are less significant if we consider the uncertainties and upper limits. Similarly, no trend between the color excess and other fitting properties can be seen when we take the uncertainties into account (Figure \ref{propmontage}). As mentioned in the last part of section \ref{sec:select}, IRAC photometry of group III LAE (\#23098) may be contaminated. By removing the LAE from group III, the derived color excesses are $0.14-0.30$ mag; any correlations between the dust extinction and magnitudes or other properties cannot be seen. The presence of some amount of color excess indicates that LAEs at this redshift are not free from dust, which is consistent with what \citet{finkel2007, finkel2008, finkel2009} found for LAEs at $z\sim4.5$. Many authors \citep{neufeld1991,hansen2006, finkel2009} proposed that the presence of dust in clumpy clouds can enhance the \lya~EW of a galaxy. To test this hypothesis, we estimated the clumpiness parameter ($q$): $q\tau$ is \lya~line opacity where $\tau$ is an optical depth for the continuum \citep{finkel2008}. If $q<1$, the \lya~EW is enhanced on the hypothesis that \lya~photons suffer less extinction than the continuum photons and the clumpy cloud model would be supported. We calculate $q$ from the ratio of observed and intrinsic \lya~luminosities. The observed \lya~luminosity is derived from the observed \lya~EW and the continuum flux density, while the intrinsic one is calculated from the \ha~luminosity by assuming case B recombination. Except for \#12766 and \#23098 that show upper limits on the $q$ parameter, all LAEs show $q > 1$ (Table \ref{cons0.2}) suggesting that \lya~photons suffer from dust extinction larger than the continuum. The uncertainty of $q$ value is large; nevertheless, the $q$ values are all near $1-2$ suggesting that although the \lya~emission appears to be more attenuated than the continuum, the difference is not so much. 
In section \ref{sec:select}, however, we mentioned that the observed \lya~EWs may be underestimated if the \lya~emission line does not fall into the central part of $NB711$ filter and are further underestimated due to the non-detection in $I_c$ band for some LAEs. Underestimation of the \lya~EW results in the overestimation of the $q$ parameter. Accordingly, the objects without spectroscopic confirmation, i.e., \#30702 and \#22962, could have $q$ values less than those shown in Table \ref{cons0.2}, if their \lya~line did not fall into the central part of the $NB711$ filter. The spectroscopic observations are desirable to investigate the real \lya~EWs and $q$ values of the LAEs. 

The best-fitting star formation rates (SFRs) of LAEs range from 55 $M_{\odot}$yr$^{-1}$ to 209 $M_{\odot}$yr$^{-1}$. The median value of the derived SFRs is 132 \Msun yr$^{-1}$. The typical error on SFR for each object is $\sim0.9$ dex. These SFRs are much higher than those of LAEs from other studies \citep[e.g.,][]{gawiser2007, nilsson2007}. This is probably due to the selection effect. Our SED fitting sample has to be bright enough to be detected in either ground-based broadbands or IRAC bands or both. Thus the selected LAEs are more luminous which results in the higher SFRs. The relation between the SFR and other fitting properties are shown in Figure \ref{propmontage}(c), (e), and (f). Looking only at the best-fitting values, we see that SFR seems to show correlation with stellar mass and dust extinction. However, the correlations are not significant when we consider the uncertainties or exclude group III LAE. 

Since some of the LAEs show very young ages, an effect of nebular continuum emission may not be negligible. We examine the effects of nebular emission on the SED fitting results and describe more details in Appendix B.  Briefly, we found that stellar mass estimation of all LAEs is robust ($\pm0.04$ dex). Age, color excess, and SFR for group I LAEs are also robust ($\pm0.24$ dex, $\pm0.02$ mag, and $\pm0.10$ dex, respectively).  For group II and III LAEs, age, color, and SFR do not agree well with each other, but agree within their large uncertainties. \ha~emission accounts for the largest expected nebular line emission to be detected from our LAEs. Including \ha~emission to the model spectrum when we performed SED fitting in the last section is likely to be the reason why adding the other nebular emissions give the similar results. 

\section{Comparisons to LBGs at $z\sim5$}\label{sec:compareLBG}
In this paper, we aim at comparing the derived stellar populations of LAEs at $z\sim5$ to those of LBGs at the same redshift and in the same field by \cite{yabe}. \citet{yabe} used LBGs at $z\sim5$ with $\zmag < 26.5$ mag selected by \citet{iwata2007}; they used the same data set and the same $V-$dropout criteria as those used in this study. 
In addition, the SED models and the fitting technique are also the same. In order to make a fair comparison, we used only LAEs whose $\zmag_{total}$ magnitudes are brighter than 26.5 mag. Because part of the LBGs are spectroscopically confirmed with the mean redshift of $z\sim4.7$ \citep{ando2007, kajino2009}, taking the same limit on \zmag~magnitude is equivalent to taking the same limit on rest-frame UV luminosity. There are six LAEs with $\zmag_{total} < 26.5$ mag in the IRAC isolated sample. Two of them are the group I LAEs; one is in group II (\#12766). The others are group IV LAEs for which we could not perform SED fitting. Therefore, we used three LAEs from group I and II for the comparison. Among these three LAEs, one (\#35915) appears in \cite{iwata2007} catalog and also in \cite{yabe}. Though we used the same set of broadband data as \cite{iwata2007} to select the sample, the images in our case were smoothed to have the same seeing size as that of the $NB711$ image before making photometry (section \ref{subsub:photosubaru}). Accordingly, magnitudes measured from the images before and after smoothing are not necessarily the same. 

\begin{figure*}[t]
\centering
\begin{tabular}{cc}
\includegraphics[trim=3mm 17mm 0mm 15mm, clip, width=6cm, angle=-90]{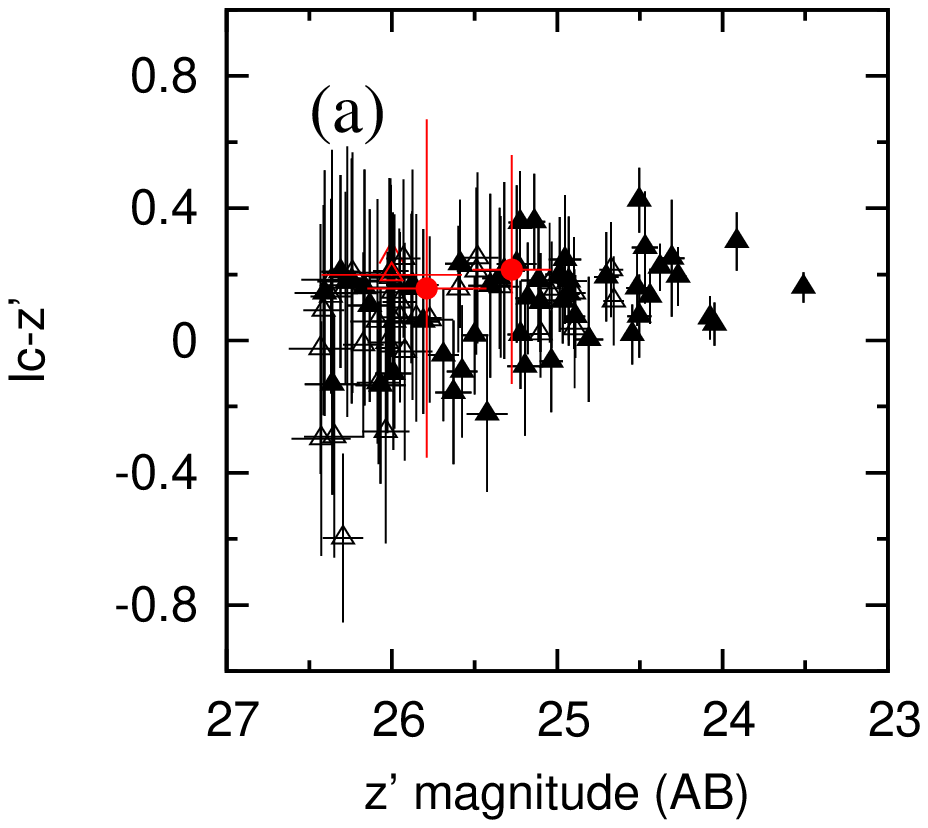} & 
\includegraphics[trim=3mm 17mm 0mm 15mm, clip, width=6cm, angle=-90]{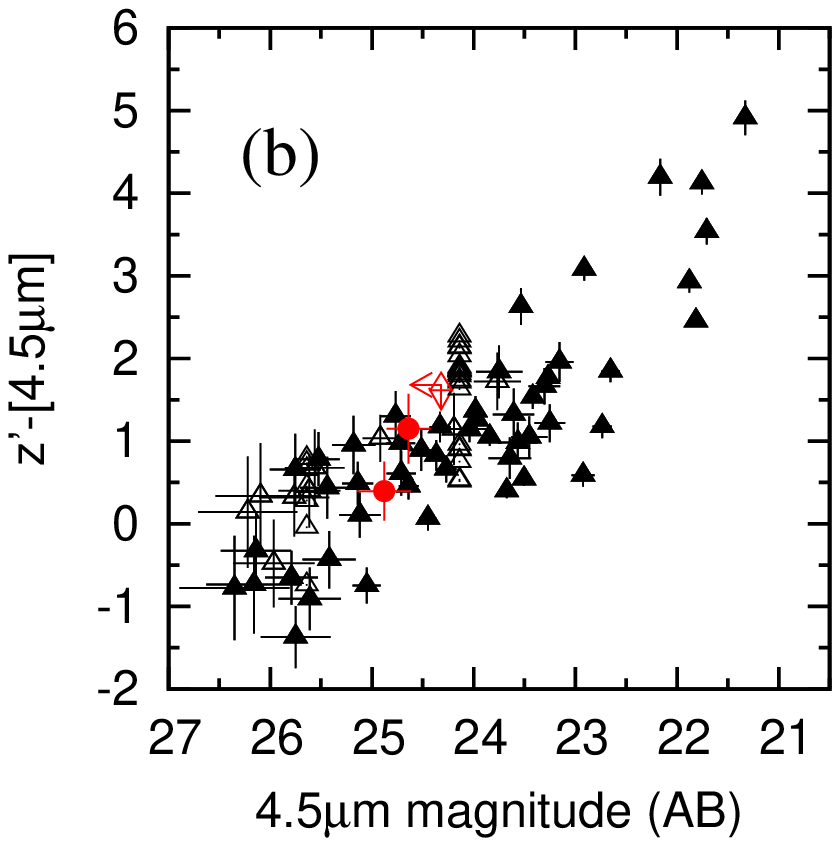}
\end{tabular}
\caption{ (a) $I_c-\zmag$ versus \zmag~magnitude diagram; (b) $\zmag-\rm{[4.5\micron]}$ versus 4.5\micron~magnitude diagram. Group I LAEs and \#12766 from group II LAEs are indicated by red filled circles and red open triangle, respectively. LBGs in category 1 are plotted in filled triangles; those in categories 2 and 3 are plotted with black open triangles. Arrows show the limits in magnitudes or colors. Errors of LBGs are taken from \cite{yabe}.}
\label{colormag}
\end{figure*}

From 617 LBGs at $z\sim5$, \citep{iwata2007}, \citet{yabe} selected 170 LBGs by eye inspection as isolated objects both in $\zmag-$band and in IRAC images. The 170 LBGs are divided into four categories according to the detection in 3.6\micron~and 4.5\micron~filters. The LBGs that are detected in both 3.6\micron~and 4.5\micron~images are in category 1, while those only detected in either 3.6\micron~or 4.5\micron~are put in category 2 or 3, respectively. The rest that are detected neither in 3.6\micron~nor 4.5\micron~images are in category 4. We used categories 1, 2, and 3 LBGs in the comparison. \cite{yabe} performed the SED fitting for 64 LBGs in category 1. For categories 2 and 3 LBGs, we refitted them by using an upper limit either in 4.5\micron~or 3.6\micron~band, respectively. 17 LBGs are best fitted with the stellar ages larger than the cosmic age at $z\sim5$ ($\sim1.2$ Gyr). The reason for the large ages is considered to be the assumption of constant star formation history or low-z interlopers (Yabe et al. 2009; Sawicki \& Yee 1998; Papovich et al. 2001). The derived ages are generally older than those derived by assuming other star formation histories. We thus excluded these objects in the following comparison. We eventually have 88 LBGs from categories 1, 2, and 3 to compare with our three LAEs from groups I and II. It should be noted that since the $NB711$ filter covers only a small fraction of the redshift range, LAEs that are not in this redshift range are expected among the LBG sample. 

Before going to the comparison between stellar populations of LAEs and those of LBGs, we try to figure out what their photometry can tell us without SED fitting. Another thing which should be kept in mind is that three LAEs which are used to compare with the LBGs in this section (i.e., group I LAEs and \#12766, hereafter "compared LAE sample") are not necessarily good representatives for all LAEs at $z=4.86$. As shown in Figure \ref{bias}, the compared LAE sample (grey histogram) is relatively bright and has comparatively red UV color as compared to the whole LAE sample. Figure \ref{colormag} shows the color-magnitude diagrams of the compared LAE sample and the LBGs at $z\sim5$. No clear trend is seen between \zmag~magnitude and $I_c-\zmag$ color (Figure \ref{colormag}(a)). The figure shows that the LAEs occupy the faint part of \zmag-magnitude distribution and the red part of $I_c-\zmag$ color distribution of the LBGs. Distribution of these LAEs at the faint part of UV luminosity among LBGs may support the deficiency of strong \lya~emission among bright LBGs claimed by \cite{ando2006}, though the sample size is only three. Figure \ref{colormag}(b) shows the color-magnitude diagram of $\zmag-\rm{[4.5\micron]}$ color and 4.5\micron~magnitude. The $\zmag-\rm{[4.5\micron]}$ colors of LBGs spread over a wide range from $-1$ to 5 mag. 
The LAEs are on the distributions of LBGs, but they locate in the slightly fainter part in 4.5\micron-band magnitude and the relatively bluer part of the color distributions than the LBGs. 
 
\begin{figure*}[ht]
  \begin{center}
    \begin{tabular}{cc}
\includegraphics[trim = 0mm 19mm 1mm 21mm, width = 6cm, angle = -90]{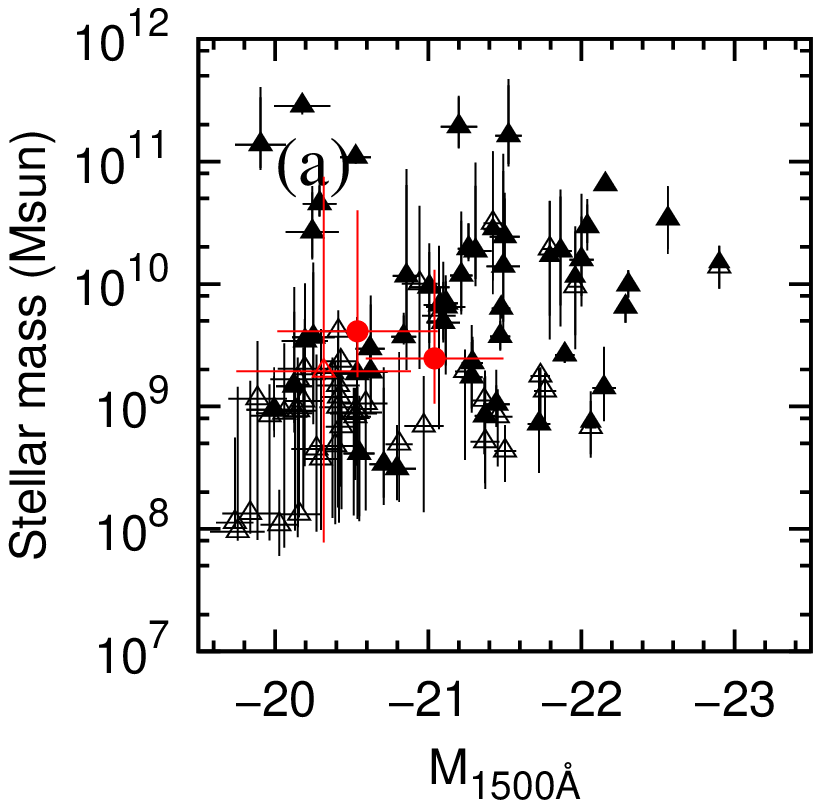} &
\includegraphics[trim = 0mm 19mm 1mm 21mm, width = 6cm, angle = -90]{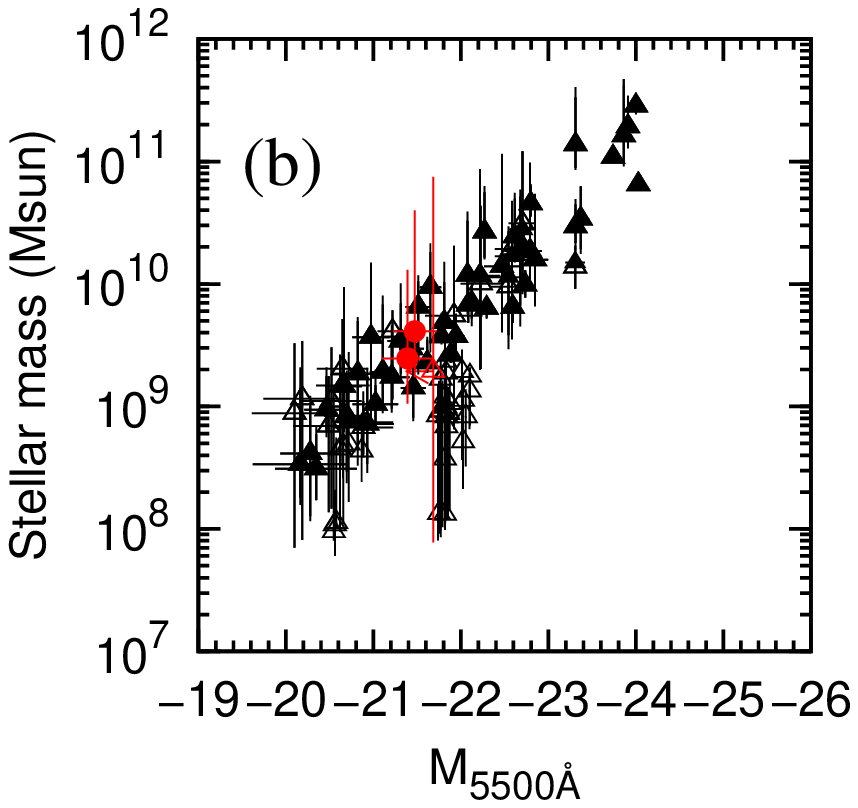} \\
\includegraphics[trim = 0mm 19mm 1mm 21mm, width = 6cm, angle = -90]{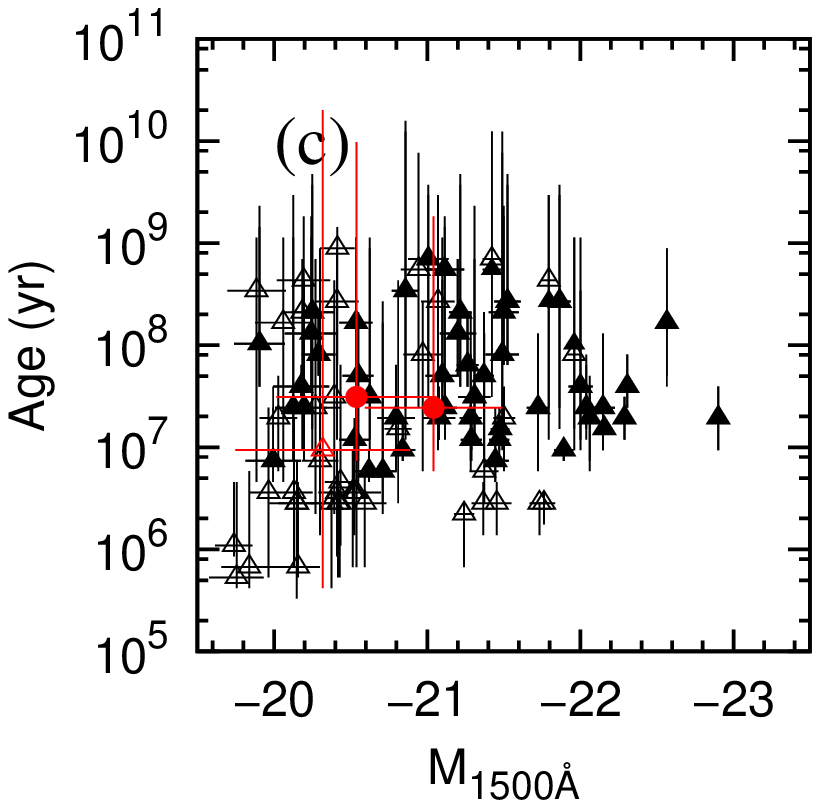} &
\includegraphics[trim = 0mm 19mm 1mm 21mm, width = 6cm, angle = -90]{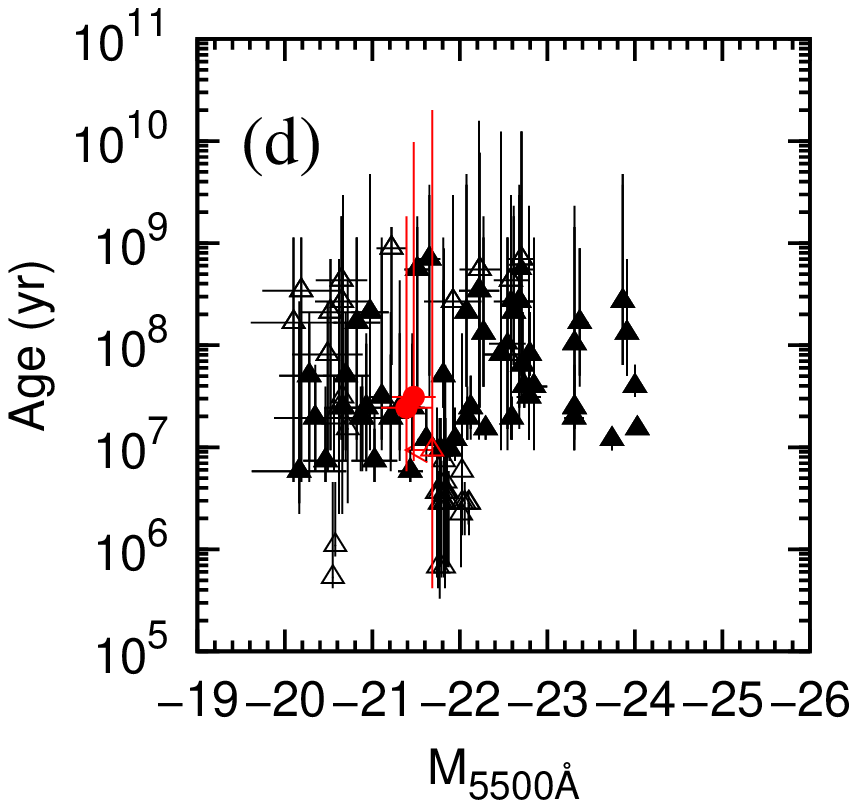} \\
\includegraphics[trim = 0mm 19mm 1mm 21mm, width = 6cm, angle = -90]{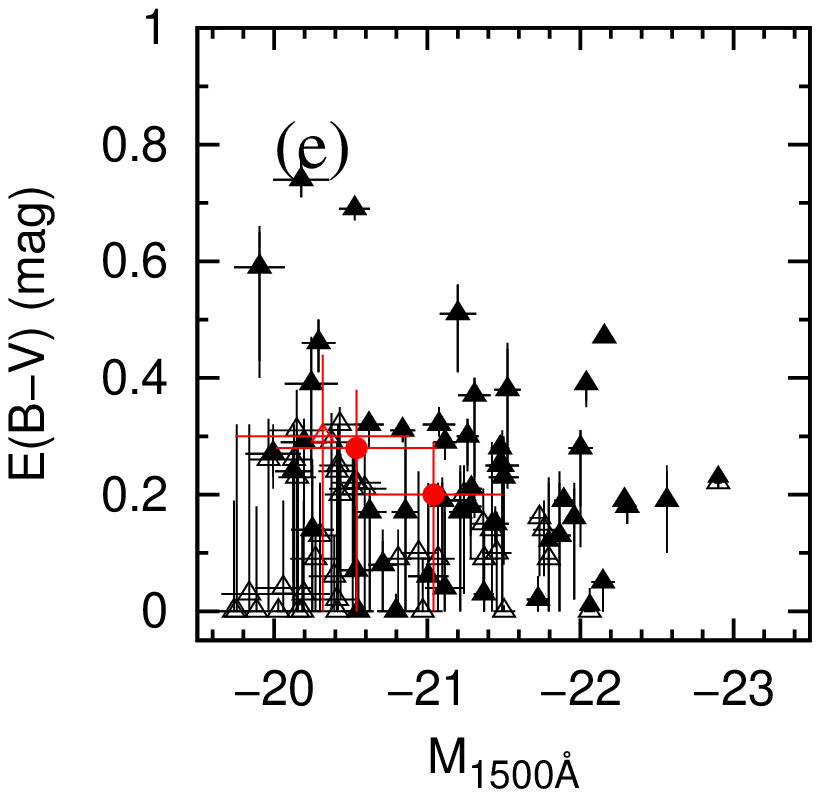} &
\includegraphics[trim = 0mm 19mm 1mm 21mm, width = 6cm, angle = -90]{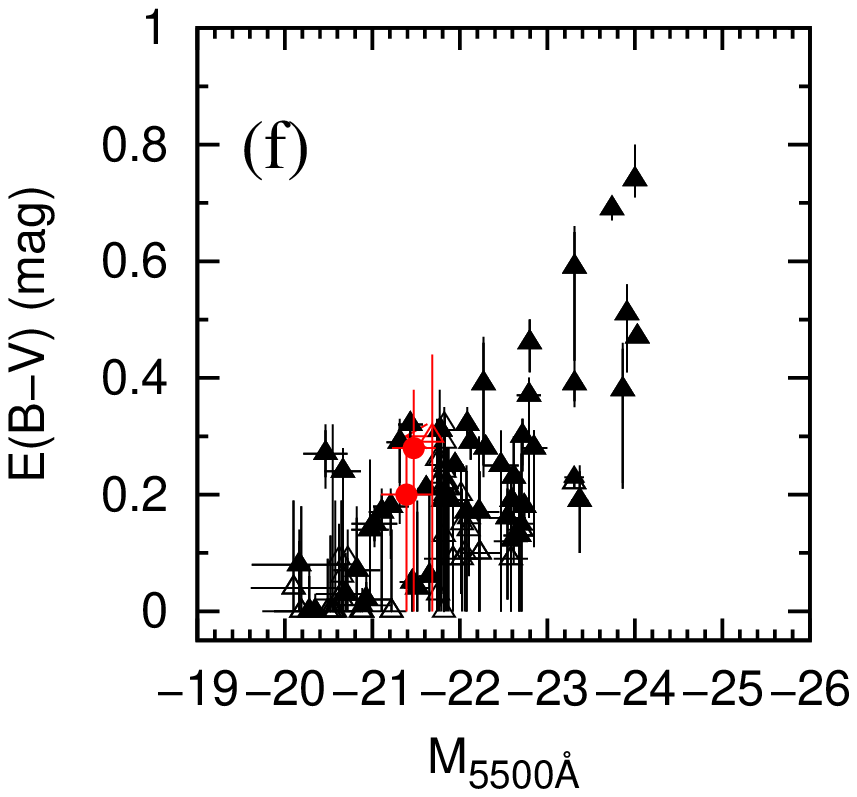} \\
\includegraphics[trim = 0mm 19mm 1mm 21mm, width = 6cm, angle = -90]{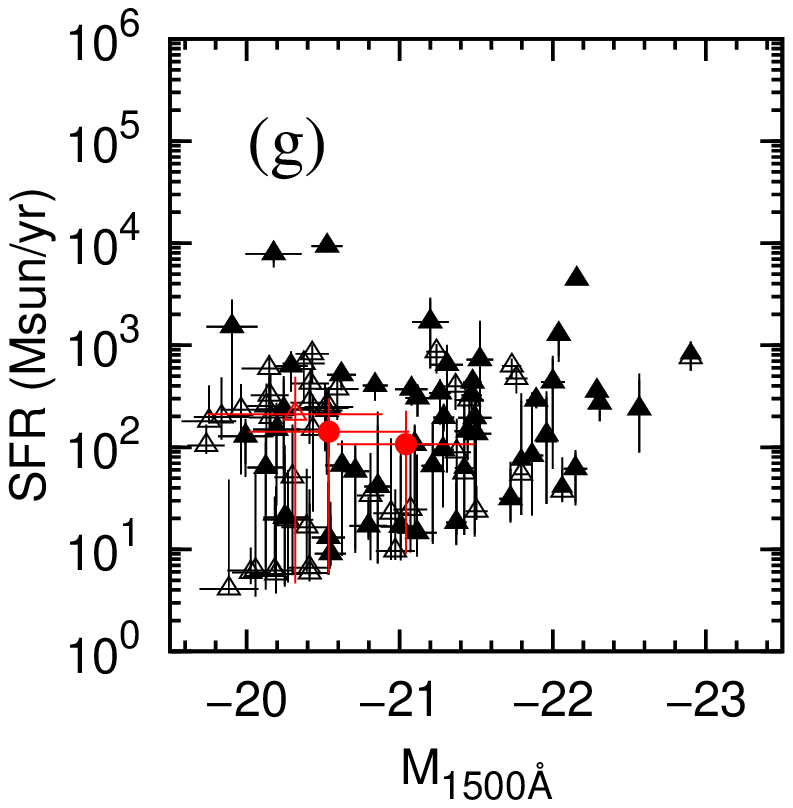} &
\includegraphics[trim = 0mm 19mm 1mm 21mm, width = 6cm, angle = -90]{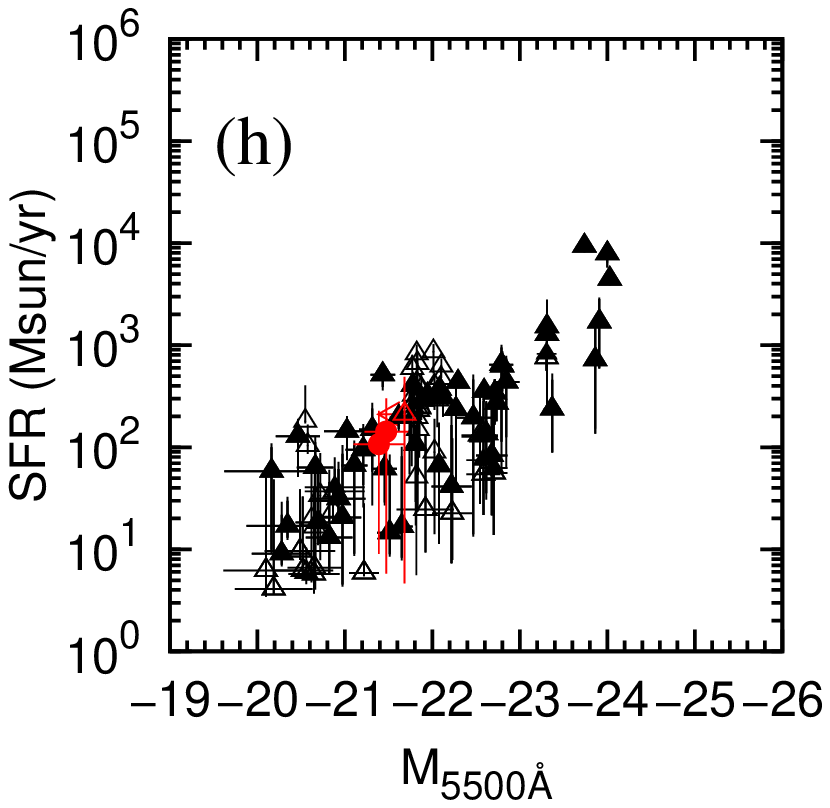} \\
   \end{tabular}
    \caption{Output parameters of SED fitting against the UV and optical absolute magnitudes. Symbols are the same as those in Figure \ref{colormag}.}
    \label{comparelbg}
  \end{center}
\end{figure*}

Output parameters of SED fitting are plotted against the rest-frame UV and optical luminosities in Figure \ref{comparelbg}. The top panels show the comparisons between the derived stellar masses of LAEs and those of LBGs. The stellar masses of the compared LAE sample range from $10^9 \Msun$ to $10^{10}\Msun$. In contrast, LBGs distribute in a larger range from $10^8\Msun$ to $10^{11}\Msun$. It is seen in the figure that the LAEs locate in the region where LBGs distribute and in the almost middle part of LBGs' distribution. At the same bin of rest-frame UV or optical luminosity, LAEs seem to have comparable masses to LBGs. 

Showing no relation between ages and luminosities, Figures \ref{comparelbg}(c) and \ref{comparelbg}(d) indicate that the LAEs again lie on 
the same distribution of LBGs. The median ages are similar; they are 25 Myr and 22 Myr for LAEs and LBGs, respectively. However, age uncertainties of the LAEs cover the whole range of age distribution of the LBGs. It is difficult to draw any conclusion.

Figures \ref{comparelbg}(e) and \ref{comparelbg}(f) show the plots of the derived color excesses against UV and optical absolute magnitudes, respectively. Although the dust extinction of the LAEs is not zero, they seem to lie at the relatively lower region of the color excess distribution of LBGs. At the fixed rest-frame UV or optical luminosity, there seems to be no difference in $E(B-V)$, even if we take the uncertainties into account. Figure \ref{comparelbg}(e) shows that there is no relation between the dust extinction and the rest-frame UV luminosity. In contrast, Figure \ref{comparelbg}(f) seems to show a correlation of LBGs between the color excesses and the optical magnitudes. Though we can not state any correlation for only 3 LAEs, the LAEs still lie on the correlation of LBGs. 

The SFRs are plotted against rest-frame UV and optical absolute magnitudes in Figures \ref{comparelbg}(g) and \ref{comparelbg}(h), respectively. Both figures show a correlation of SFRs of the LBGs with UV (but weak) and optical luminosity. It is seen from the figures that the LAEs locate in the lower part of the LBGs' distribution. A median SFR of the LAEs is 132 \Msun yr$^{-1}$, while it is 187 \Msun yr$^{-1}$ for LBGs. Note that the LAEs used in the comparison are biased toward the brightest sample among the whole LAE sample (section \ref{sec:bias}), which probably results in the higher SFRs of the LAEs. In conclusion, no significant differences in the stellar properties between LAEs and LBGs are seen at the same luminosity. 

\begin{figure*}[htbp]
\centering
\begin{tabular}{cc}
\includegraphics[trim = 0mm 12mm 0mm 26mm, width = 6cm, angle = -90]{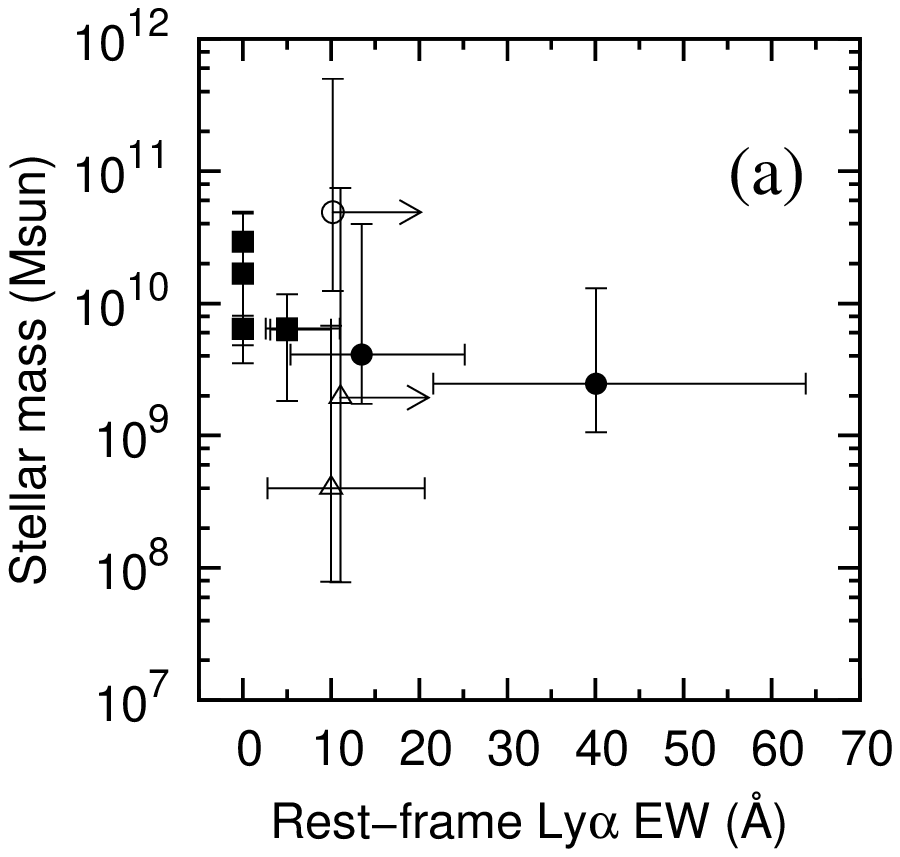} &
\includegraphics[trim = 0mm 12mm 0mm 26mm, width = 6cm, angle = -90]{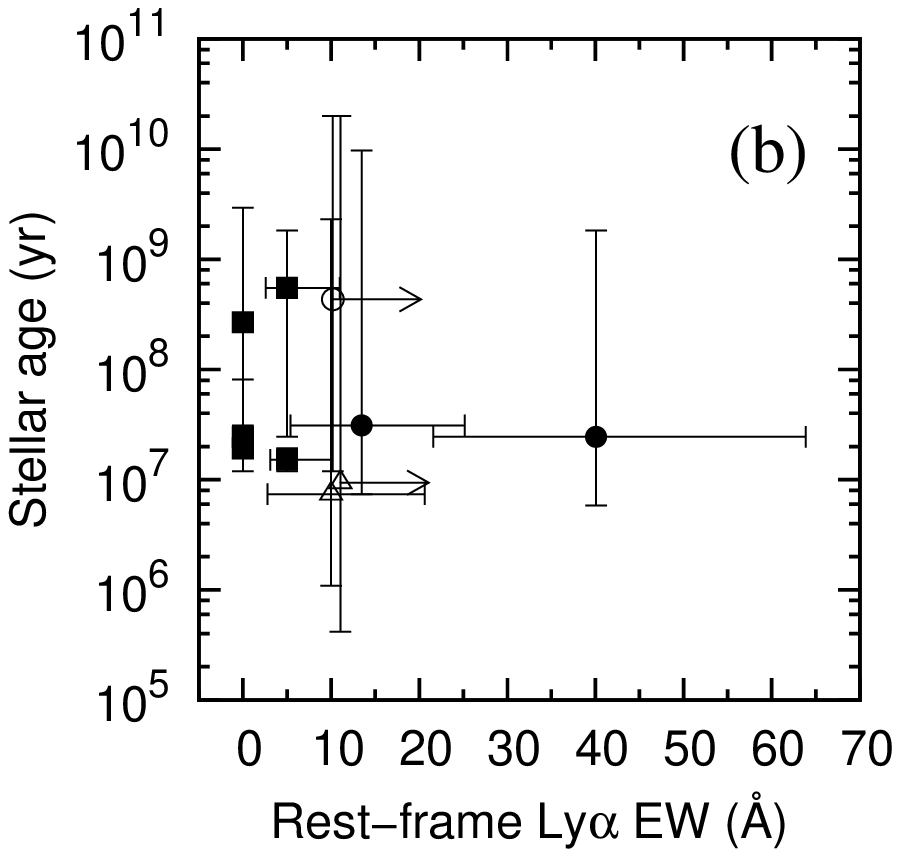} \\
\includegraphics[trim = 0mm 12mm 0mm 26mm, width = 6cm, angle = -90]{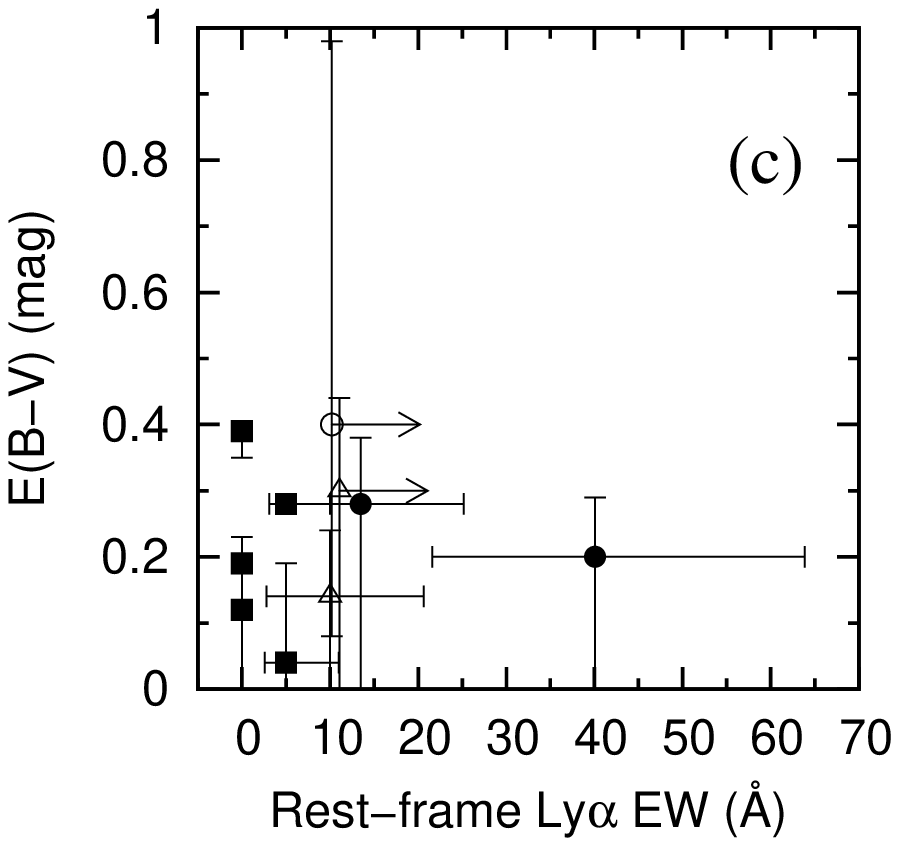} &
\includegraphics[trim = 0mm 12mm 0mm 26mm, width = 6cm, angle = -90]{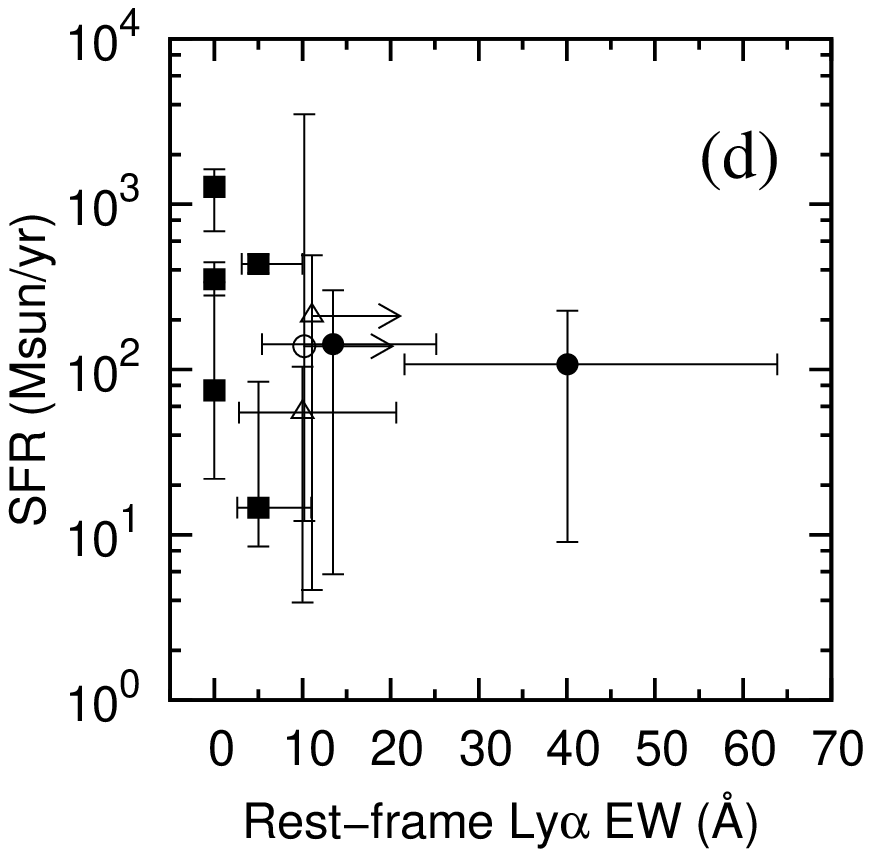} \\
\end{tabular}
\caption{Output parameters from SED fitting for group I--III LAEs versus rest-frame \lya~equivalent widths. Filled circles, open triangles, and open circles represent group I, group II, and group III LAEs, respectively. LBGs that have spectroscopic \lya~EWs are also shown in filled squares. }
\label{ewprop}
\end{figure*}

A physical reason that makes an LAE an emitter is still unclear. In order to investigate whether or not the difference between LBGs and LAEs has any dependence on the equivalent widths (EWs) of \lya~emission, we plotted the output parameters of SED fitting against the rest-frame \lya~EWs in Figure \ref{ewprop}. Group I--III LAEs are plotted with five LBGs which have spectroscopic \lya~EWs \citep{ando2004, kajino2009}. Arrows in the figures represent the lower limits on EWs of the LAEs; $I_c$ magnitudes of them are fainter than $2\sigma$ limiting magnitude. Note that, except for one spectroscopically confirmed LAE, the \lya~EWs of LAEs are likely to be lower limits if their \lya~emission lines do not fall into the center of $NB711$ band. According to the figure, we cannot find any significant correlation between the stellar properties and the rest-frame \lya~EWs. Recently, \citet{kornei2009} studied these relations for $z\sim3$ LBGs with spectra. They found that the stellar mass does not correlate with the \lya~EW, while large \lya~EW is seen in older, lower SFR, and less dusty LBGs. 
Such trends are difficult to isolate with only three LAEs. A larger sample at $z\sim5$ is required to see clearer relations if any and to be definitive. 

\section{Summary}\label{sec:conclusion}

In this paper, we studied the stellar properties of Lyman alpha emitters (LAEs) at $z=4.86$ using SED fitting. By narrowband and broadband observations by Suprime-Cam on Subaru telescope, 24 LAEs were selected in the area of $\sim508.5$ arcmin$^2$ around GOODS-N field. In addition to the optical photometry, we obtained the mid-infrared photometry from data taken by IRAC on Spitzer space telescope in the GOODS-N field as well as the surrounding area in order to cover most part of Subaru area. We selected 12 LAEs that are isolated from the neighboring objects. We performed SED fitting of five LAEs which are detected above $2\sigma$ magnitude limits in more than 2 bands. Selecting those five LAEs could introduce a bias toward the bright red galaxies. Model SEDs are built by assuming the constant star formation history with fixed metallicity of 0.2\Zsun, the Salpeter IMF ranging from 0.1\Msun~to 100 \Msun, and the extinction law of Calzetti et al. (2000). The derived stellar masses of the LAEs range from $10^8$ to $10^{10}$ $M_{\odot}$ with the median value of $2.5\times10^9\Msun$. The derived ages cover wide range from 7.4 Myr to 437 Myr with the median value of 25 Myr. 
The color excess are between $0.1-0.4$ mag, indicative of the presence of some amount of dust. Star formation rates (SFRs) are in the range of $55-209$ $M_{\odot}$yr$^{-1}$. The median color excess and SFRs are 0.27 mag and $132 M_{\odot}$yr$^{-1}$, respectively. The high SFRs are probably due to the selection effect; we selected the LAEs that are bright enough to be detected in rest-frame UV and optical bands, which results in selecting the LAEs with the higher SFRs.  We investigate the correlations between the stellar properties derived by SED fitting and the photometric properties of LAEs and found no significant correlation due to both small size of the sample and the large uncertainty on fitting results. 

The main objective of this study is to compare LAEs to LBGs at the same redshift. LBGs were selected by V-dropout criteria (Iwata et al.2007). Their stellar populations were derived by Yabe et al. (2009). Because those LBGs are selected from the same set of data and stellar population was derived by the same SED fitting model, we can make a fair comparison between their stellar populations. We compared three LAEs to 88 LBGs down to the same UV luminosity limit. These three LAEs are the brightest and reddest ones among the whole LAE sample in this study. The comparisons of SED-fitting parameters show that LAEs locate in the region where LBGs distribute; the physical properties of LAEs and LBGs occupy similar parameter spaces. At the same rest-frame UV or optical luminosity, there is no difference of stellar properties between LAEs and LBGs. In order to figure out properties that control the \lya~EWs, we plotted the output parameters against the rest-frame \lya~EWs ranging from 0 to 40 \AA. We could not find any significant correlation between them. A larger number of sample is needed to see any correlation of \lya~EW if exists. 

We thank the referee for their valuable comments which improved the paper. This work is supported by the Grant-in-Aid for Scientific Research on Priority Areas (19047003) and by the Grant-in-Aid for Global COE program "The Next Generation of Physics, Spun from Universality and Emergence" from Ministry of Education, Culture, Sports, Science, and Technology of Japan.

\appendix

\section{Effects of nebular emission}
In order to investigate the effects of nebular emission on the synthesis models, we used the stellar population synthesis model by Fioc \& Rocca-Volmerange (1997, 1999; hereafter PEGASE). The PEGASE can add nebular emission (continuum and emission lines) to the model spectrum. The hydrogen emission lines are computed from the number of Lyman continuum photons by assuming case B recombination. Other emission lines are calculated from the observed ratios to H$\beta$ for typical local starbursts \citep{fioc1997}. Other prescriptions are the same as those used in the case of BC03 model described in the main text. We used Salpeter IMF (1995) with lower and upper mass cutoffs of 0.1 and 100 \Msun, respectively, adopted metallicity of 0.2\Zsun, and assumed the constant star formation history. We re-scaled the age step of PEGASE into the logarithmic scale as we did for BC03 model. The effect of dust attenuation is taken into account by using Calzetti extinction law \citep{calzetti2000}. $E(B-V)$ varies from 0.0 mag to 0.8 mag by 0.01 mag step. The attenuation due to IGM is followed by \citet{madau1995}. The redshift is fixed at $z=4.86$ except for \#35915 which has a spectroscopic redshift (section \ref{sec:sedfit}). 

\begin{figure}[htbp]
\centering
\includegraphics[clip, angle=-90, width=10cm]{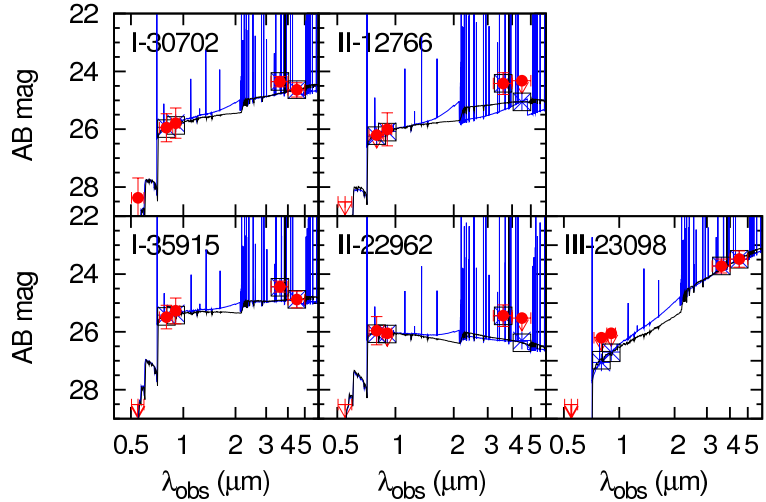}\\
\caption[Plot of observed SEDs and best fit model spectra]
{Observed SEDs and the best-fitting models for LAEs in group I (left panel), group II (middle panel), and group III (right panel). In each panel, the observed SED is shown with filled circles; the best-fitting model SEDs are indicated by opened squares and asterisks for BC03 and PEGASE models, respectively. Black line represents the best-fitting BC03 model spectrum with \ha~emission line, while blue one represents the PEGASE model. $2\sigma$ upper limit of V-band photometry is also shown with the arrow; however, the data is not used in SED fitting. 
}
\label{allspec}
\end{figure}

\begin{figure}[ht]
  \begin{center}
    \begin{tabular}{cc}
\includegraphics[trim = 0mm 19mm 1mm 21mm, width = 5cm, angle = -90]{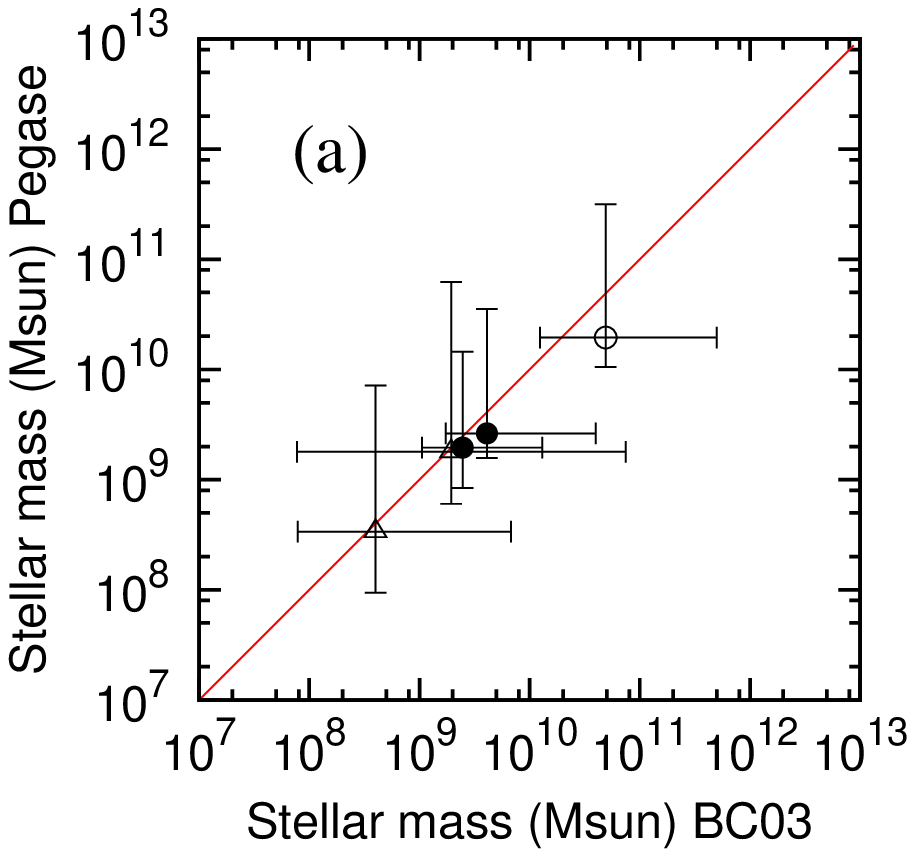} &
\includegraphics[trim = 0mm 19mm 1mm 21mm, width = 5cm, angle = -90]{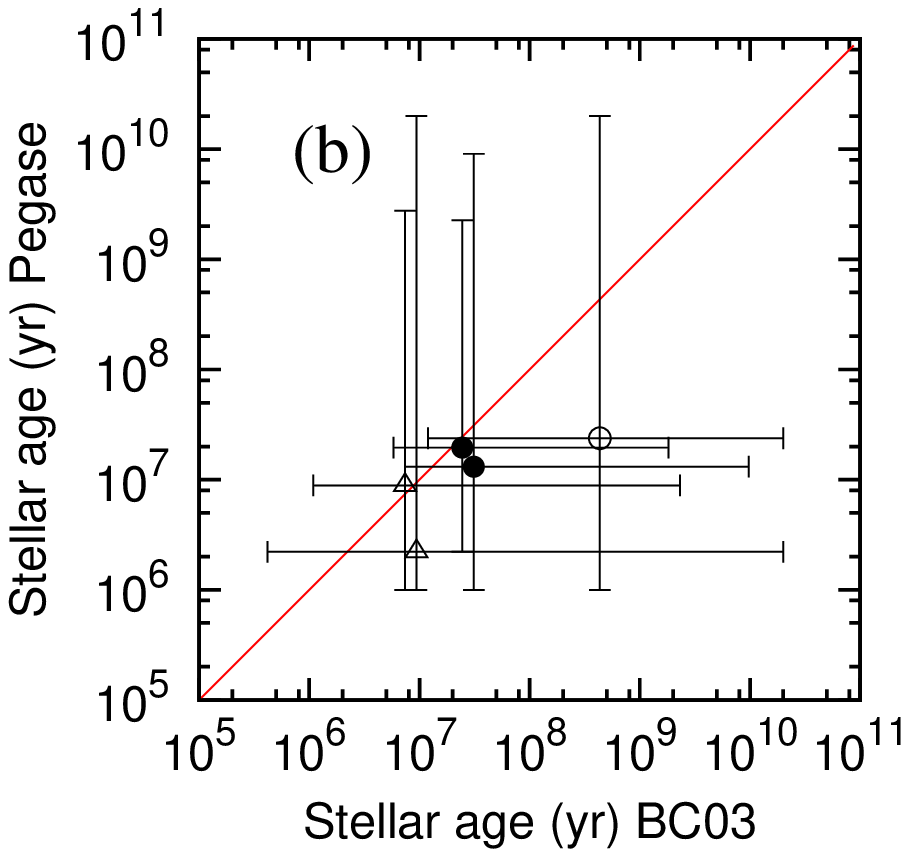} \\
\includegraphics[trim = 0mm 19mm 1mm 21mm, width = 5cm, angle = -90]{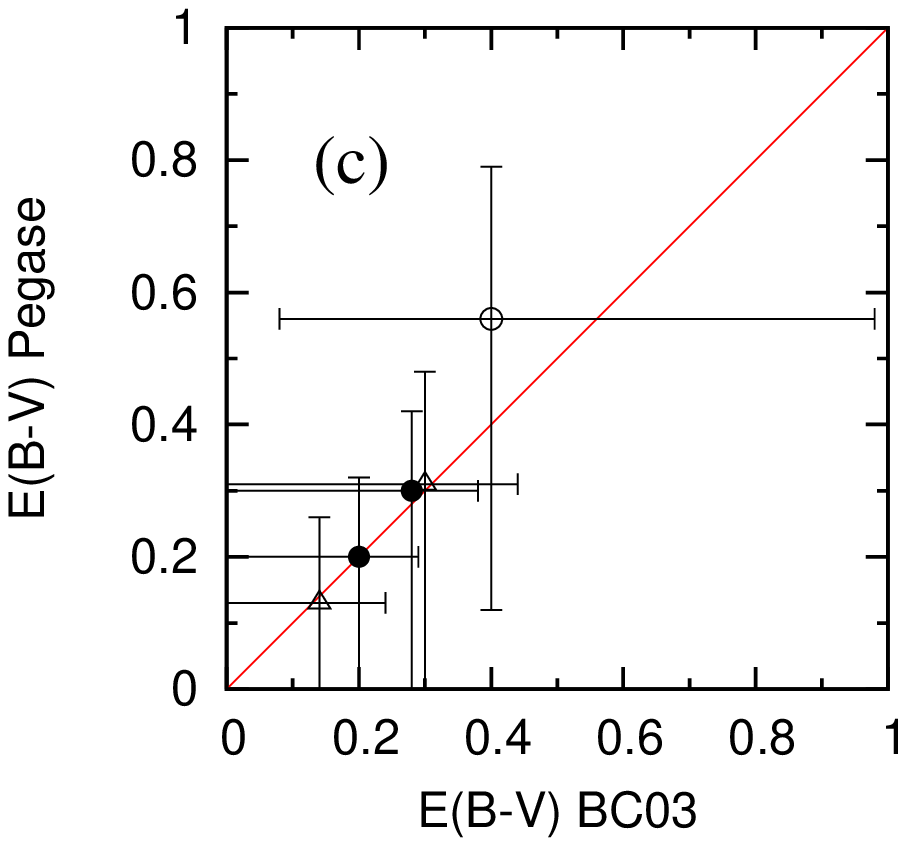} &
\includegraphics[trim = 0mm 19mm 1mm 21mm, width = 5cm, angle = -90]{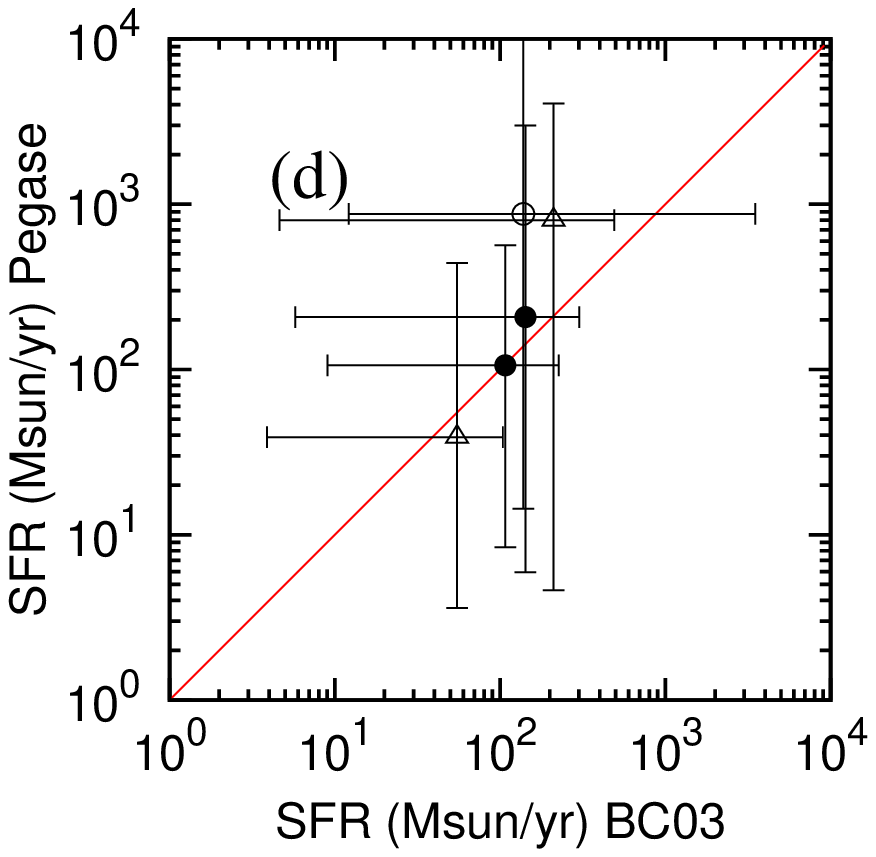} \\
   \end{tabular}
    \caption{Comparison of output parameters from BC03 and Pegase models. Filled circles, open triangles, and open circles represent group I, group II, group III LAEs, respectively.}
    \label{comparepegase}
  \end{center}
\end{figure}

The results obtained for group I--III LAEs with PEGASE are summarized in table \ref{pegasetable}. The best-fit spectra are shown as a comparison with BC03 best-fitting spectra in Figure \ref{allspec}. The comparisons of resulting parameters are presented in Figure \ref{comparepegase}. Figure \ref{comparepegase}(a) shows that the stellar masses derived by both models are well in agreement with each other. The stellar masses obtained by the PEGASE model are averagely smaller than those by the BC03 model by about 0.04 dex. As seen in Figure \ref{comparepegase}(b), the stellar ages of most LAEs derived by PEGASE are younger than those derived from BC03 models. Inclusion of nebular emission averagely decreases the ages of all LAEs by 0.46 dex. However, due to the large uncertainties in determining the stellar ages, the derived ages seem to be in agreement within the uncertainty range. Figure \ref{comparepegase}(c) shows that the color excesses derived by PEGASE are comparable to those by BC03 model. The average difference is 0.04 mag except for the group III LAE. As seen in Figure \ref{comparepegase}(d), the SFRs derived by both models seem to be comparable for most of the LAEs. However, the rather large differences are seen for the group III LAE and one in group II (\#12766). SFRs increase averagely about 0.28 dex when nebular emission is included. 

\begin{deluxetable}{c c c ccc c}
\tabletypesize{\footnotesize}
\tablewidth{0pt}
\tablecaption{The best fit results for PEGASE model \label{pegasetable}}
\tablehead{
\colhead{ID} & \colhead{Field} & \colhead{log[Mass]} & \colhead{log[Age]} & \colhead{E(B-V)} & \colhead{log[SFR]} & \colhead{$\chi^2_\nu$} \\
\colhead{} & \colhead{} & \colhead{(\Msun)} & \colhead{(yr)} & \colhead{(mag)} & \colhead{(\Msun yr$^{-1}$)} & \colhead{} 
}
\startdata
\multicolumn{6}{l} {Group I}&\\
\hline
30702 & GOODS-N & $9.42_{-0.22}^{+1.13}$ & $7.12_{-1.12}^{+2.84}$ & $0.30_{-0.30}^{+0.12}$ & $2.32_{-1.54}^{+1.16}$ & $0.001$ \\
35915 & GOODS-N & $9.29_{-0.37}^{+0.87}$ & $7.29_{-0.95}^{+2.06}$ & $0.20_{-0.20}^{+0.12}$ & $2.03_{-1.10}^{+0.73}$ & $0.05$ \\
\hline
\multicolumn{6}{l} {Group II} &\\
\hline
12766 & GOODS-FF & $9.25_{-0.47}^{+1.54}$ & $6.34_{-0.34}^{+3.96}$ & $0.31_{-0.31}^{+0.17}$ & $2.90_{-2.24}^{+0.71}$ & $44.88$\\
22962 & GOODS-N & $8.53_{-0.56}^{+1.33}$ & $6.95_{-0.95}^{+2.49}$ & $0.13_{-0.13}^{+0.13}$ & $1.59_{-1.03}^{+1.05}$ & $46.64$ \\
\hline
\multicolumn{6}{l} {Group III}&\\
\hline
23098 & GOODS-FF & $10.29_{-0.27}^{+1.21}$ & $7.38_{-1.38}^{+2.92}$ & $0.56_{-0.44}^{+0.23}$ & $2.94_{-1.78}^{+1.53}$ & $47.05$\\
\enddata
\end{deluxetable}
\end{document}